\documentclass[preprint,12pt]{elsarticle}

\usepackage{amssymb}
\usepackage{amsmath}
\usepackage{url}
\usepackage{subfigure}
\usepackage{graphicx}
\usepackage{xcolor}
\usepackage{amsthm}
\usepackage{amsmath}
\usepackage[normalem]{ulem}
\usepackage{multirow}
\usepackage{makecell}
\journal{Medical Image Analysis}

\begin{document}

\begin{frontmatter}



\title{MoRe-3DGSMR: Motion-resolved reconstruction framework for free-breathing pulmonary MRI based on 3D Gaussian representation}


\author{Tengya Peng$^1$, Ruyi Zha$^2$, Qing Zou$^{3,4,5}$} %% Author name

% \affiliation{organization={},%Department and Organization
\address{$^1$Department of Biomedical Engineering, University of Texas Southwestern Medical Center, Dallas, TX, USA\\
$^2$Australian National University, Canberra, Australia\\
$^3$Department of Pediatrics, University of Texas Southwestern Medical Center, Dallas, TX, USA\\
$^4$Department of Radiology, University of Texas Southwestern Medical Center, Dallas, TX, USA\\
$^5$Advanced Imaging Research Center, University of Texas Southwestern Medical Center, Dallas, TX, USA}

%% Abstract
\begin{abstract}
This study presents a self-supervised, motion-resolved reconstruction framework for high-resolution, free-breathing pulmonary magnetic resonance imaging (MRI) using a three-dimensional Gaussian representation (3DGS). The proposed method leverages 3DGS to address the challenges of motion-resolved 3D isotropic pulmonary MRI reconstruction by enabling smooth interpolation across voxels for continuous spatial representation across different respiratory motion states. Pulmonary MRI data are acquired using a golden-angle radial sampling trajectory, with respiratory motion signals extracted from the center of k-space in each radial spoke. Based on the DC self-gated respiratory signal, the k-space data are sorted into multiple respiratory motion states. A 3DGS framework is then applied to reconstruct a reference image volume from the first motion state. Subsequently, a learnable low-rank sparse motion model is trained to estimate deformation vector fields (DVFs) from the canonical state to the other motion states, which are then used to generate images for the remaining motion states by spatially transforming the reference volume. The proposed reconstruction pipeline is evaluated on six subjects and benchmarked against three state-of-the-art MRI motion-resolved reconstruction methods. The experimental findings demonstrate that the proposed framework effectively reconstructs high-resolution, motion-resolved pulmonary MR images. Compared with existing approaches, it achieves superior image quality, as reflected by higher signal-to-noise ratio and contrast-to-noise ratio. The proposed self-supervised 3DGS-based reconstruction method enables pulmonary MRI to be resolved into distinct respiratory states with isotropic spatial resolution, highlighting its promise for clinical pulmonary MRI.
\end{abstract}

%% Keywords

\end{frontmatter}



\section{Introduction}
Magnetic resonance imaging (MRI) holds significant promise for evaluating pulmonary diseases, owing to its superior soft tissue contrast and ability to delineate structural details of the lungs \cite{hatabu2020expanding,ohno2022overview,sharma2022quantification}. However, pulmonary MRI is technically challenging due to the inherently short $T_2^{*}$ relaxation time of lung tissue and the presence of respiratory motion artifacts. To address the rapid signal decay associated with short $T_2^{*}$, specialized pulse sequences such as ultrashort echo time (UTE) \cite{sanchez2023detection,larson2024lung} and zero echo time (ZTE) \cite{wang2025assessment,ufuk2024comparing} have been developed. These sequences use optimized excitation schemes and fast readout trajectories to achieve sub-millisecond echo times (TE), thereby enhancing signal-to-noise ratio (SNR) in lung imaging.

Most UTE and ZTE implementations adopt non-Cartesian k-space trajectories, such as three-dimensional (3D) radial sampling, to maximize sampling efficiency and improve robustness to motion \cite{gandhi2024comparison,yu2021free,wu20234d}. In particular, center-out 3D radial acquisitions with golden-angle rotation have become increasingly popular because they facilitate rapid data acquisition and high-quality image reconstruction \cite{johnson2013optimized,zhu2021optimizing}. Moreover, the retrospective sorting capability of 3D radial sampling provides additional flexibility for motion-compensated reconstruction algorithms \cite{zou2023time,zou2024motion}. Despite these advances, high-resolution 3D UTE/ZTE acquisitions typically require several minutes of scan time, during which respiratory motion can degrade image quality through motion-induced artifacts, even when non-Cartesian sampling is used. To mitigate motion artifacts, respiratory signals can be acquired using external devices such as respiratory belts, which monitor abdominal expansion during respiration. However, signals from external devices may be unreliable in some cases \cite{montazeri2021design}. Alternatively, self-navigation methods exploit the repeatedly sampled center of k-space, also known as the direct current (DC) signal, to estimate respiratory motion directly from MRI data. This approach is particularly compatible with center-out UTE sequences and enables motion tracking without the need for external hardware.

Respiratory motion signals extracted using self-navigation techniques enable retrospective sorting of k-space data into discrete respiratory phases, thereby facilitating motion-resolved image reconstruction. Traditional reconstruction strategies primarily rely on compressed sensing (CS) methodologies \cite{feng2014golden}, wherein initial reconstructions -- typically obtained via inverse Fourier transforms -- are iteratively refined through low-rank or sparsity-based regularization. These methods aim to suppress noise and artifacts while preserving critical anatomical structures. One notable example is XD-GRASP (eXtra-Dimensional Golden-angle RAdial Sparse Parallel imaging) \cite{feng2016xd}, which integrates compressed sensing with parallel imaging using multi-dimensional encoding and temporal gradient sparsity to accelerate reconstruction and reduce sampling requirements. More recently, ExtremeMRI \cite{ong2020extreme} proposed a multiscale low-rank (MSLR) factorization model, representing the spatiotemporal data matrix as a sum of block-wise low-rank matrices across multiple scales, and optimized the model using stochastic gradient methods. With the emergence of deep learning, newer approaches have unrolled the iterative optimization steps of CS into trainable neural network frameworks \cite{murray2024movienet,jafari2023graspnet}; however, such methods usually require extensive training data, which is typically not feasible in the setting of 3D non-Cartesian MRI. An alternative paradigm, typically implemented in a self-supervised fashion, directly treats the MR volume representation as a learnable parameter. In this formulation, model parameters are optimized by backpropagating through the Fourier transform to minimize the discrepancy between predicted and acquired k-space data. For instance, MoCo-SToRM \cite{zou2022dynamic} introduced a method in which a learnable tensor encodes the dynamic MR volume, and a neural decoder is used to predict deformation vector fields (DVFs) \cite{li2018efficient}, enabling the transformation of a reference image into various motion states.

Three-dimensional Gaussian representation (3DGS) has recently emerged as a powerful technique for scene representation and novel view synthesis \cite{kerbl20233d}, owing to its ability to model complex 3D structures with high fidelity through the smooth and continuous nature of Gaussian distribution functions \cite{bersillon2001use}. In this framework, these distribution functions --hereafter referred to as Gaussian points -- serve as spatially localized primitives that encode both geometric and radiometric properties. Each Gaussian point is characterized by attributes such as position, anisotropic scale, and orientation, which collectively influence its contribution to the overall rendered scene. By capturing spatial correlations within a probabilistic representation, 3DGS facilitates efficient and accurate modeling of three-dimensional structures. The strengths of 3DGS have recently been extended to medical imaging applications. Zha et al. \cite{zha2024r} successfully employed 3D Gaussian splatting for cone-beam computed tomography (CBCT) reconstruction. Peng et al. \cite{peng2025three} further extended 3DGS to complex MRI reconstruction, adapting it to accommodate the complex-valued nature of MRI data rather than the real-valued setting typical of X-ray and CBCT modalities, and demonstrated its ability to address the undersampling problem in Cartesian MRI. In this work, we further adapt and extend 3DGS for isotropic non-Cartesian motion-resolved pulmonary MRI reconstruction.

Motion models play a vital role in motion-related MRI reconstruction, motivating the development of compact and learnable motion representations that can capture the sparsity and spatial continuity of deformation while predicting coherent DVFs across motion states or time. Prior studies have shown that PCA-based lung motion models provide an effective low-dimensional representation of respiratory deformation \cite{li20113d}. By exploiting the strong spatiotemporal correlations of pulmonary motion, lung DVFs can be approximated by a few motion bases and their corresponding coefficients. In particular, two principal components are theoretically sufficient for regular respiratory motion \cite{li2011pca}, and clinical studies further suggest that two PCA coefficients are often adequate for accurate motion modeling. These findings indicate that pulmonary motion lies in a compact low-rank subspace, motivating learnable DVF models based on a small number of spatial bases and temporal coefficients for compact and subject-specific motion representation \cite{shao2025real,huang2025digs,chen2025multi}. Chen et al. \cite{chen2025multi} in particular employed a deep image prior (DIP)-based low-rank CNN model for DVF generation. Such compact CNN parameterizations in DIP have been shown to capture low-level structural information and provide an implicit prior that regularizes the representation \cite{liang2025analysis}. These advances and innovations motivate us to propose a learnable low-rank sparse motion model specifically designed for motion-resolved MRI reconstruction. Notably, in the motion-resolved scheme, respiratory states are naturally ordered, and the corresponding motion evolves across states in a temporally coherent and approximately monotonic manner. Our motion model explicitly incorporates this property, enabling more compact and physiologically consistent DVF representation across states.

The primary aim of this study is to integrate respiratory motion information into the 3DGS framework for motion-resolved reconstruction of pulmonary MRI. To achieve this, respiratory motion signals were first estimated from the center of k-space data, enabling retrospective binning of the k-space into a fixed number of discrete motion states. The first motion state, designated as the reference state, was reconstructed using 3DGS. To enhance the convergence and stability of the reconstruction process, a novel multi-resolution scaled initialization strategy was introduced. Simultaneously, a learnable low-rank sparse motion model was employed to estimate the DVFs from the first state to the subsequent states, which were then used to generate the remaining motion states by spatially transforming the reference state. For each motion state, a non-uniform fast Fourier transform (NUFFT) was performed to simulate corresponding k-space data, which was then compared against the actual binned k-space data for that respiratory phase to enforce data consistency during training.

The key contributions and innovations of this study are summarized as follows: 1) Introduced 3D Gaussian representation for 4D MRI representation in a motion-resolved framework. 2) Demonstrated the effectiveness of 3DGS under the non-uniform Fourier transform and extended it to non-Cartesian and non-uniform sampling scenarios. 3) Developed a multi-scale resolution scaling method to initialize 3D Gaussian points at different scales from low- to high-resolution MRI k-space, which accelerated convergence and improved reconstruction performance. 4) Proposed a learnable low-rank sparse model specific to the motion-resolved scenario.
 
\section{Methods}

For clarity, throughout this section, we use \textit{state} to denote a reconstruction bin obtained from retrospective k-space sorting, and \textit{phase} to denote named respiratory stages (e.g., end-inspiration, mid-inspiration, mid-expiration, and end-expiration).

\subsection{4D MRI reconstruction from radial sampling}

The 4D MRI reconstruction (a.k.a motion-resolved reconstruction) of free-breathing, ungated 3D pulmonary MRI acquired using center-out radial sampling begins with the estimation of respiratory motion from the central k-space data, commonly referred to as the DC signal. In this study, respiratory motion estimation was performed following the methodology outlined in \cite{zhu2020iterative}. Specifically, the multi-coil k-space data were first combined using the adaptive navigator strategy \cite{zhang2016robust}. The resulting DC signals were subsequently processed using a low-pass filter with a cutoff frequency between 0.5 and 1 Hz to attenuate high-frequency noise components.

Following motion signal estimation, the k-space data were retrospectively sorted into $N_{state}$ respiratory motion states, ranging from end-expiration to end-inspiration. The proposed motion-resolved reconstruction framework is formulated as a learning task that jointly estimates an image representation $\mathbf{X} = [x_\mathbf{j}]$  at the reference motion state, where $x_{\mathbf{j}}$ denotes the complex-valued intensity at the $j$-th voxel $(x_j,y_j,z_j)$ and the reference motion state is defined as the first respiratory phase in this work, together with a low-rank sparse motion model $D_{\theta}$ that outputs DVF corresponding to each respiratory phase, denoted by $\mathbf{u}_{t}$  for $t = 1,\cdots, N_{state}$. The reference image representation $\mathbf{X}$ can then be optimized by solving a data fidelity-driven minimization problem that enforces consistency with the acquired motion-binned k-space measurements:  

\begin{equation}\label{key}
\begin{gathered}
\arg \min_{\mathbf{X}, \theta} \sum_{t}^{N_{state}}  \mathbf{W} \cdot \left \|\mathcal{F}\cdot\mathbf{C} \cdot warp(\mathbf{X},\mathbf{u}_{t})-\mathbf{m}_{t}  \right \|+\lambda _{tv}\cdot TV(\mathbf{X}),
\\
\mathbf{u}_{t}=D_{\theta}(t),\quad \mathbf{u}_{t} \in \mathbb{R}^{256 \times 256 \times 256 \times 3}
\\
\quad \mathbf{X} \in \mathbb{C}^{256 \times 256 \times 256},\quad \mathbf{W} \in \mathbb{R}^{256 \times 256 \times 256},\quad \mathbf{C} \in \mathbb{C}^{256 \times 256 \times 256}.
\end{gathered}
\end{equation} 

Note the above equation only accounts for the spatial part of the loss. The motion-resolved consistency loss is depicted in Section 2.4. Here, the warp function is implemented using a spatial transformer. $\mathcal{F}$ is the NUFFT operator and $\mathbf{C}$ is the coil sensitivity maps. $\mathbf{m}_{t} = [\mathbf{m}_1, \cdots, \mathbf{m}_N]^T$ represents the sorted k-space data for the $N_{state}$ motion states. $\theta$ denotes the decoder network parameters.  $TV$ denotes the total variation regularization term \cite{rodriguez2013total} computed on the gridded reference-state image, and $\lambda _{tv}$ is the corresponding regularization weight.

\subsection{3D Gaussian representation for 3D MRI}

In this work, we represent the reference image representation $\mathbf{X}$ in equation \eqref{key} using 3DGS. Specifically, we represent each $x_{\mathbf{j}}$ as a linear combination of a set of Gaussian points. Mathematically speaking,  
\begin{equation}\label{image_model}
x_{\mathbf{j}} = \sum_{i=1}^{M}G_i^3(\mathbf{j} | \mathbf{\rho}_i, \mathbf{p}_i, \mathbf{\Sigma_i}),
\end{equation} 
where $M$ is the number of Gaussian points to represent each voxel and
 
\begin{equation}\label{key2}
\begin{gathered}
G_i^3(\mathbf{j} | \rho_i, \mathbf{p}_i, \mathbf{\Sigma}_i) = 
\rho_i\cdot\exp\left(-\frac{1}{2}(\mathbf{j}-\mathbf{p}_i)^T\mathbf{\Sigma}_{i}^{-1}(\mathbf{j}-\mathbf{p}_i)\right), \\
\mathbf{j}=(x_j,y_j,z_j),\quad \mathbf{p}_i \in \mathbb{R}^{3},\quad  \rho_i \in \mathbf{C},\quad \mathbf{\Sigma}_i \in \mathbb{R}^{3\times3}.
\end{gathered}
\end{equation}

Here, $\mathbf{p}_{i}$ is the center location of the each Gaussian point. $\rho_i$ is a complex number to represent the central density value of the Gaussian point. $\mathbf{\Sigma}_i$ defines the shape size and orientation of the Gaussian point, and can be represented as 
\begin{equation}\label{key3}
\mathbf{\Sigma}_i = \mathbf{R}_i\mathbf{S}_i\mathbf{S}_i^T\mathbf{R}_i^T,
\end{equation} 
where $R_i$ and $S_i$ represents the rotation and scaling matrix respectively. The rotation matrix is represented by pairs of iso-rotations characterized by quaternions. The scaling matrix quantifies the size of Gaussian points in three directions.

In \eqref{image_model}, each Gaussian point is characterized by a set of learnable parameters, including its spatial scale, anisotropic orientation, and central density value. Each Gaussian point exerts influence over a localized region within the volume, specifically within a radius of $3\sigma$, where $\sigma$ denotes the length of the principal axis of the corresponding Gaussian ellipsoid. This $3\sigma$ range covers the vast majority of the Gaussian mass and is therefore commonly adopted in prior studies \cite{zha2024r,peng2025three}. This localized influence allows the 3DGS to compactly model complex 3D structures while maintaining accurate spatial continuity and smoothness. Furthermore, Equation \eqref{image_model} defines the voxelization process, which maps the continuous 3D Gaussian point cloud into a discretized voxel grid, thereby enabling integration with conventional volumetric processing pipelines. 
%This voxelization process is deterministic.

 The initialization of 3DGS in this work follows the procedure described in \cite{peng2025three}. Specifically, an inverse NUFFT is first applied to the binned k-space data of the reference motion state to obtain a coarse volumetric representation in the image domain. A threshold set to 0.05 of the maximum magnitude is then applied to the complex-valued image to separate foreground candidate grid points from the background. In \cite{peng2025three}, a random initialization strategy was adopted for the low-to-mid under-samled ratios, in which $M$ grid points were randomly sampled from these candidate points and used as the initial Gaussian centers. In the present work, we further extend this scheme by proposing a multi-resolution initialization strategy, which is described in detail in the following section. The Gaussian scales are initialized according to the three nearest neighboring Gaussian points, and each scale is set to three times the average distance to these neighbors. The Gaussian points are assumed to have zero rotation at initialization, and their rotations are subsequently learned during optimization.

Adaptive control \cite{charatan2024pixelsplat} serves as a key component in the original 3DGS framework, facilitating the dynamic densification of Gaussian points. This process enhances the representational capacity and structural fidelity of the reconstruction by incrementally introducing additional Gaussian points. However, in this work, an ablation study demonstrated that initializing the reconstruction with the maximum number of Gaussian points, while omitting the adaptive densification process, resulted in improved training efficiency without compromising reconstruction quality. Consequently, adaptive control was omitted in this work.

%Our previous work revealed the best performance was achieved when initialization started with sparse points, followed by densification through a coarse-to-fine training process. In this study, we applied the same adaptive control strategy, including splitting and cloning techniques.

\subsection{Multi-resolution Initialization Strategy}

Accurate initialization plays a critical role in ensuring effective performance in 3D Gaussian-based reconstruction frameworks \cite{jung2024relaxing}, which involves generating the initial 3D Gaussian points based on the inverse NUFFT of the motion-sorted radial k-space data corresponding to the first respiratory state.  Random initialization remains a commonly adopted approach in existing literature \cite{kerbl20233d,zha2024r}, where the spatial positions of the Gaussian points are randomly sampled within the volume. Their initial scales are computed as the average distance to their three nearest neighboring Gaussian points. As a result, the initialized Gaussian points exhibit variability in both spatial location and scale.

It has been demonstrated in previous research \cite{peng2025three} that equal-space initialization yields superior performance in the context of reconstruction from undersampled data, primarily due to the enhanced capability of uniformly distributed, small-scale Gaussian points to capture high-frequency image components. In this initialization strategy, Gaussian points are initialized with approximately equal inter-point spacing and consistent scale parameters. This uniform configuration facilitates faster convergence towards accurate representation. 

\begin{figure}[tb!]
\centering
   \includegraphics[width=0.6\textwidth]{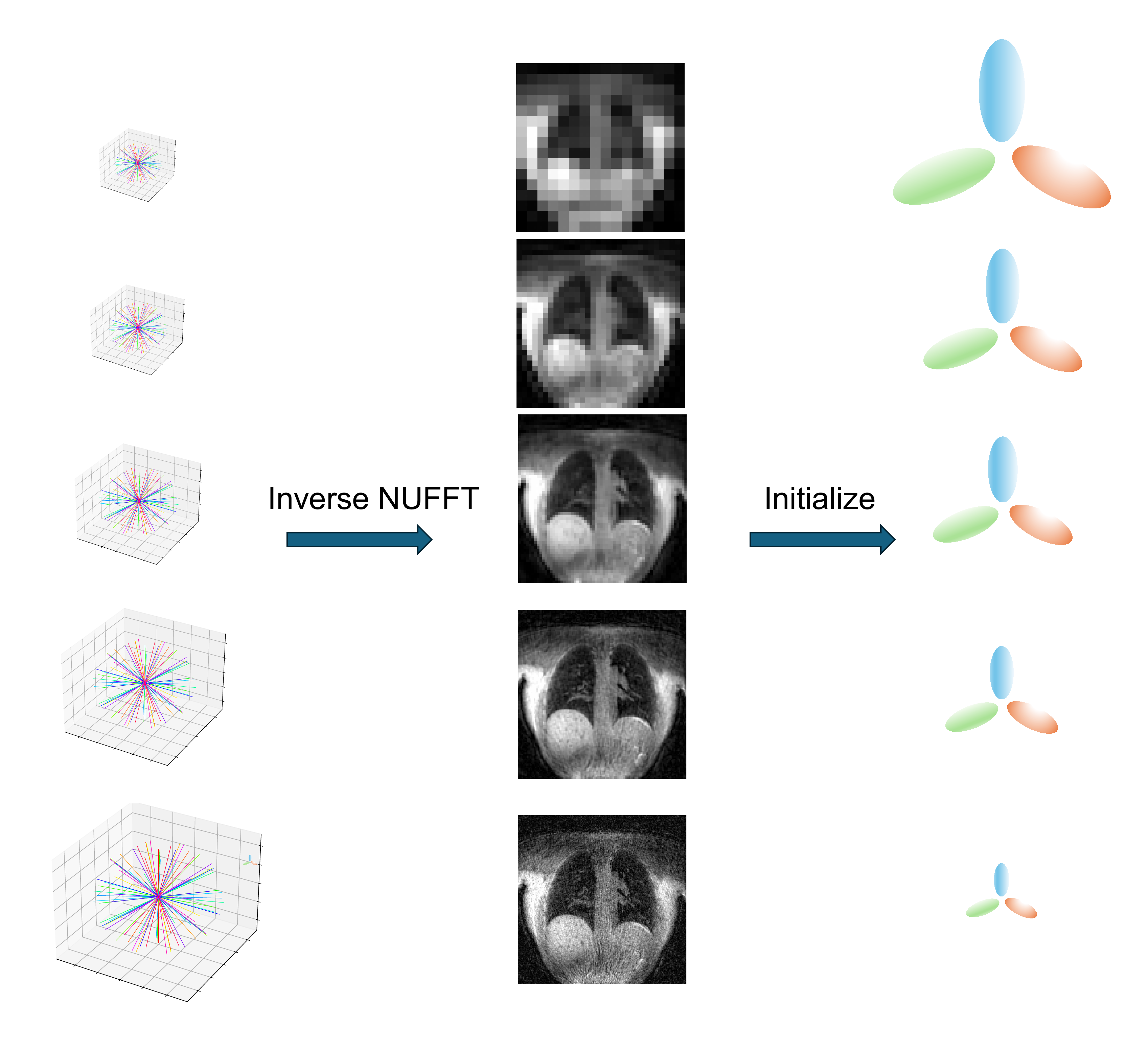}

  \caption{Illustration of the multi-resolution initialization strategy. Gaussian points are initialized at multiple spatial scales using outputs from inverse Fourier transforms computed at progressively increasing levels of k-space coverage, forming a hierarchical (pyramidal) architecture. Each resolution level corresponds to a subset of the k-space data, with higher levels incorporating a greater number of samples per radial spoke to progressively recover higher-frequency information. The Gaussian points initialized at each level are summed, and then simultaneously refined during the training process.}
  \label{multires_fig}
\end{figure}

In this study, a multi-resolution initialization strategy is proposed to enhance the representation capability of 3DGS for motion-resolved MRI reconstruction, due to the nature of radial k-space sampling and the amount of k-space in each motion states. In contrast to the existing coarse-to-fine training paradigm that trains Gaussian points with densification progressively from low resolution to high resolution, the proposed multi-resolution initialization strategy initializes Gaussian points based on the frequency domain information, progressing from coarse-to-fine, and then the initialized Gaussian points with different sizes are trained simultaneously. Specifically, inverse NUFFT operations are applied to frequency-filtered k-space data across multiple resolution levels, forming a resolution pyramid. At each level, Gaussian points are initialized using an equal-space initialization. Larger Gaussian points are allocated to low-frequency components to capture coarse anatomical structures, while smaller Gaussian points are assigned to high-frequency components to preserve fine-grained details. The numbers of Gaussian points assigned to different resolution levels are also designed in a cascade manner. Specifically, the $l$-th level is assigned $4^{l}\cdot n$ Gaussian points, so that the total number of Gaussian points satisfies: 
\begin{equation}\label{multi_resolution_N}
\sum_{l=0}^{L=5}4^{l}\cdot n =M,
\end{equation} 
where $L$ represents the total number of resolution levels, $l$ denotes the current level, and $M$ is the total number of Gaussian points. From the finest level to the coarsest level, the k-space data are progressively undersampled along each radial spoke by retaining only every other sampling point. As a result, the sampling density along the radial direction is halved and the spacing between adjacent samples on each spoke is doubled level by level. This initialization method is depicted in Figure \ref{multires_fig}. By enabling simultaneous training of Gaussian points across multiple resolution levels during training, this approach eliminates the need for a separate coarse-to-fine training phase with adaptive densification.

%\begin{figure}[htbp!]
%\centering
%   \includegraphics[width=0.3\textwidth]{figs/equal_space.png}
%  \caption{Equal-space initialization for 3DGS. The Gaussian points are initialized with equal space and same scales.}
%\end{figure}

A learnable low-rank sparse motion model for motion-resolved MRI

In motion-resolved reconstruction, k-space data are retrospectively sorted into $N_{state}$ uniformly sorted respiratory phases based on the estimated motion signal, covering the full respiratory cycle from end-inspiration to end-expiration. In the proposed framework, motion correction is achieved through a learnable low-rank sparse motion model $D_{\theta}$ that estimates DVFs warpping the first motion state, which is the reference motion states, to the rest of the motion states. The overall structure of $D_{\theta}$ is shown in Figure~\ref{pipeline_fig}. A DIP-based CNN is employed to estimate the motion basis at a coarse spatial scale using a volume downsampled by a factor of 4 relative to the original resolution. The motion basis is parameterized over $N_{state}$ states, each associated with two components, $DVF_{\alpha_{i}}$ and $DVF_{\beta_{i}}$, motivated by prior PCA studies indicating that pulmonary respiratory motion can often be represented accurately using two principal components. These two components are then linearly combined with two learnable time-varying coefficients, $\alpha$ and $\beta$, to form the final coarse DVF as \begin{equation}
DVF_{state_i}=\alpha_{i} \cdot DVF_{\alpha_{i}} + \beta_{i} \cdot DVF_{\beta_{i}}.
\end{equation} The final fine-scale DVFs are upsampled by bilinear interpolation to match the MRI volume resolution. To better adapt the learnable low-rank sparse motion model to the motion-resolved framework, $\alpha_{i}$ was parameterized by the median value of the centered DC self-gated signal for each state. As the respiratory states are naturally ordered by breathing amplitude, this parameterization enforces $\alpha_{i}$ to vary monotonically across states. Meanwhile, $\beta_{i}$ is kept freely learnable to capture minor residual motion variations across states.

\begin{table}[t]
\centering
\small
\caption{Structure of DIP-based CNN}
\label{tab:decoder64}
\begin{tabular}{|p{3.2cm}|c|c|c|c|c|}
\hline
Layer & Filters & Kernel & Stride & Pad & Output \\
\hline
Input latent & 16 & -- & -- & -- & \(8\times8\times8\) \\
\hline
\makecell[l]{Conv3d + Norm \\ + LeakyReLU} & 64 & 3 & 1 & 1 & \(8\times8\times8\) \\
\hline
\makecell[l]{Conv3d + Norm \\ + LeakyReLU} & 64 & 3 & 1 & 1 & \(8\times8\times8\) \\
\hline
\makecell[l]{Upsample \\ (factor=2)} & -- & -- & -- & -- & \(16\times16\times16\) \\
\hline
\makecell[l]{Conv3d + Norm \\ + LeakyReLU} & 32 & 3 & 1 & 1 & \(16\times16\times16\) \\
\hline
\makecell[l]{Conv3d + Norm \\ + LeakyReLU} & 32 & 3 & 1 & 1 & \(16\times16\times16\) \\
\hline
\makecell[l]{Upsample \\ (factor=2)} & -- & -- & -- & -- & \(32\times32\times32\) \\
\hline
\makecell[l]{Conv3d + Norm \\ + LeakyReLU} & 16 & 3 & 1 & 1 & \(32\times32\times32\) \\
\hline
\makecell[l]{Conv3d + Norm \\ + LeakyReLU} & 16 & 3 & 1 & 1 & \(32\times32\times32\) \\
\hline
\makecell[l]{Upsample \\ (factor=2)} & -- & -- & -- & -- & \(64\times64\times64\) \\
\hline
\makecell[l]{Conv3d + Norm \\ + LeakyReLU} & 8 & 3 & 1 & 1 & \(64\times64\times64\) \\
\hline
\makecell[l]{Conv3d + Norm \\ + LeakyReLU} & 8 & 3 & 1 & 1 & \(64\times64\times64\) \\
\hline
Conv3d + Tanh & 12 & 1 & 1 & 0 & \(64\times64\times64\) \\
\hline
\end{tabular}
\label{tab:DIP-CNN}
\end{table}

The structure of the DIP-based CNN is shown in Table \ref{tab:DIP-CNN}. It provides a compact parameterization of the coarse DVF components by mapping a low-dimensional latent tensor, initialized from uniformly sampled random noise, to a high-dimensional deformation field through a small set of convolutional filters. This restricted representation implicitly constrains the solution to a low-dimensional, effectively low-rank manifold. Moreover, the coarse-to-fine upsampling and convolution operations favor smooth, spatially correlated, and low-frequency deformation patterns in the early stages, thereby naturally preserving low-resolution structural priors while suppressing noise-like or highly irregular motion. Within the overall pipeline, the DIP-based CNN is optimized in a self-supervised manner so that the learned low-dimensional DVF components are driven to match the corresponding k-space measurements.

Motion-related 4D reconstruction typically requires additional regularization along the temporal or motion-state axis to enforce consistency and smooth transitions across adjacent frames or respiratory states. In our framework, because images at the subsequent states are generated by warping the canonical state with the corresponding DVFs, we regularize the coarse DVFs using both spatial and phase-wise smoothness terms shown as
\begin{equation}\label{L_DVF}
\mathcal{L}_{DVF} = \mathcal{L}_{spatial} + \mathcal{L}_{phase}.
\end{equation} 
The spatial term penalizes first-order spatial differences to encourage locally smooth deformation within each state, whereas the phase-wise term penalizes weighted differences between DVFs across respiratory states to enforce coherent motion evolution. The weights are determined by the distances between normalized respiratory amplitudes, so that nearby states are encouraged to have more similar deformation fields. The spatial term is presented as below:

\begin{equation}\label{L_spatial}
\mathcal{L}_{\mathrm{spatial}}
=
\frac{1}{|\Omega|}
\sum_{t=1}^{N_{state}}
\sum_{\mathbf{r}\in\Omega}
\sum_{d\in\{x,y,z\}}
\left\|
\mathbf{u}_t(\mathbf{r}+\mathbf{e}_d)-\mathbf{u}_t(\mathbf{r})
\right\|_2^2.
\end{equation}  where $\mathbf{u}_t$ is the coarse DVF of a state from $N_{state}$ states, $\mathbf{r}$ denotes the spatial position, and $\mathbf{e}_d$ denotes the unit vector along the $d$-th direction.
The phase term is depicted as:

\begin{equation}
\mathcal{L}_{\mathrm{phase}}
=
\frac{
\sum_{i=1}^{N_{state}}\sum_{j=1}^{N_{state}} w_{ij}\, d^2(\mathbf{u}_i,\mathbf{u}_j)
}{
\sum_{i=1}^{N_{state}}\sum_{j=1}^{N_{state}} w_{ij} + \varepsilon
}
\end{equation} 
where $d^2(\mathbf{u}_i,\mathbf{u}_j)$ denotes the mean squared difference between the two deformation vector fields of two motion states, and $\varepsilon$ is a small positive constant added to the denominator for numerical stability. It is normalized by  $w_{ij}$ as:

\begin{equation}
w_{ij}
=
\exp\!\left(
-\frac{(s_i-s_j)^2}{2\sigma^2+\varepsilon}
\right),
\qquad
w_{ii}=0
\end{equation}

where $s$ is the median DC self-gated signal amplitude of the state.

\subsection{MoRe-3DGSMR}

\begin{figure}
\centering
   \includegraphics[width=1\textwidth]{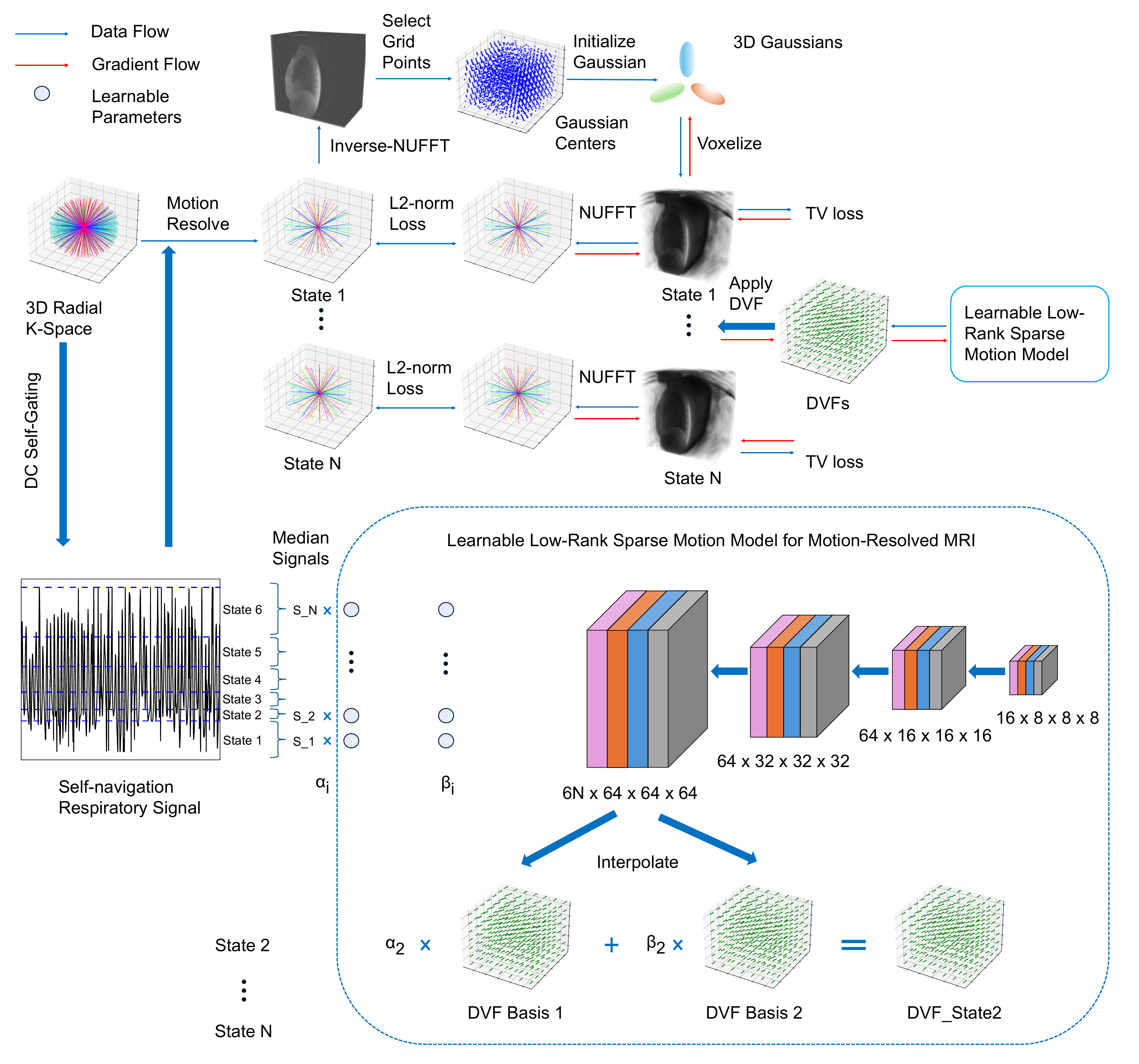}
  \caption{Overall pipeline for MoRe-3DGSMR. The framework comprises the following key components: respiratory signal prediction through DC self-gating, k-space binning, multi-resolution Gaussian points initialization, voxelization, DVFs prediction by the learnable low-rank sparse motion model, motion warping, and NUFFT operations. The estimated respiratory signal is used to retrospectively sort radial k-space spokes into discrete motion states corresponding to different respiratory phases. The k-space data from the first motion state are utilized for initializing the Gaussian points. During training, these Gaussian points are iteratively optimized and subsequently voxelized into volumetric representations.}
  \label{pipeline_fig}
\end{figure}

Integrating the aforementioned components yields the proposed motion-resolved reconstruction framework for isotropic-resolution 3D pulmonary MRI, referred to as MoRe-3DGSMR. The overall reconstruction pipeline is depicted in Figure \ref{pipeline_fig}. In this framework, the proposed 3DGS framework described above is employed to reconstruct the reference motion state (i.e., the first respiratory phase). The continuous Gaussian representation is mapped to a discrete volumetric grid through a differentiable voxelization operator. During the forward pass, each grid point accumulates contributions from neighboring Gaussian points according to their spatial weights. During the backward pass, gradients from the gridded volume are propagated back to the Gaussian parameters through the adjoint of the same weighted accumulation process, which is implemented explicitly in CUDA. For more details, please refer to \cite{peng2025three}. The images at the subsequent motion states are generated by warping the reference-state image using the DVFs predicted by the learnable low-rank sparse motion model. For each motion state, the corresponding k-space data are computed using the differentiable NUFFT operator. Model optimization is performed by minimizing the discrepancy between the computed k-space and the motion-sorted k-space measurements. Gradients of this k-space data-consistency loss are first propagated to the complex-valued voxel grid through the differentiable NUFFT operator, and then further back-propagated to the Gaussian parameters through the voxelization operator. The overall loss function used for motion-resolved reconstruction is then defined as follows: 
%  \textcolor{blue}{
% \begin{equation}
% \hspace*{-7.5cm}\mathbf{u}_t = D_{\theta}(t), \quad t=1,2,3,\ldots,N_{state}
% \label{eq:ut}
% \end{equation}
% }

\begin{equation}\label{key7}
\begin{aligned}
&\mathcal{L}(\rho_i, \mathbf{p}_i, \mathbf{\Sigma}_i,\theta) = \sum_{t}^{N_{state}} \left\| \mathbf{W} \cdot \mathcal{F} \cdot \mathbf{C} \cdot \text{warp} \left( \sum_{i=1}^{M} G_i^3 (\mathbf{j} | \rho_i, \mathbf{p}_i, \mathbf{\Sigma}_i), \mathbf{u}_t\right) - \mathbf{m}_{t}  \right\| \\
 &+\lambda_{tv} \cdot TV \left( \sum_{i=1}^{M} G_i^3 (\mathbf{j} | \rho_i, \mathbf{p}_i, \mathbf{\Sigma}_i) \right)\\ &+ \lambda_{spatial} \cdot \mathcal{L}_{spatial} +\lambda_{phase} \cdot \mathcal{L}_{phase}.
\end{aligned}
\end{equation}

The Gaussian points and the motion model are optimized simultaneously by minimizing the loss function. Once the Gaussian points and the motion model are sufficiently trained, the 3DGS is voxelized to obtain the volume representation for the first motion states. The images of subsequent motion states are obtained by applying the DVFs predicted by the motion model to the image of the first motion state.

\subsection{Dataset}

Six healthy volunteers were enrolled in this study. The data acquisition protocol was approved by the local Institutional Review Board, and written informed consent was obtained from all participants prior to imaging. All datasets were acquired using a 3D center-out radial UTE sequence on a 3T GE MRI scanner equipped with AIR coils. The imaging protocol employed 86,000 uniformly distributed golden-angle radial spokes, with a repetition time (TR) of 3.7 ms, an echo time (TE) of 0.1 ms, and 149 samples per readout, resulting in a total acquisition time of approximately 5.5 minutes. To reduce data dimensionality while preserving signal quality, a PCA-based coil compression method \cite{pedersen2009k} was applied, retaining eight virtual coils for subsequent reconstruction. The prescribed field of view $= 32 \times 32\times 32$ cm$^3$ and the reconstruction matrix size $=256\times 256\times 256$.

\subsection{Implementation details}

The proposed MoRe-3DGSMR framework was implemented using the PyTorch library, with the SigPy library employed for the NUFFT operation. All experiments were conducted on a single NVIDIA A6000 GPU, which provides 48 GB of memory. Although increasing the number of Gaussian points improves the representational capacity of the model, it also increases the computational cost and training time. Therefore, in this study, the maximum number of Gaussian points was constrained by the available GPU memory and limited to 30,000. We further conducted ablation studies on the number of Gaussian points and the coefficient of TV regularization to show the impacts. The coefficients of the spatial regularization term,
$\lambda_{spatial}$, and the phase term, $\lambda_{phase}$, are both set to 0.005. 

 Our framework employs two Adam optimizers. One is used to optimize the Gaussian parameters, while the other is used to train the learnable motion model, including the DIP-based CNN and the motion-state-specific coefficients. For the Gaussian optimizer, the learning rates are set to 0.01 for the complex-valued densities 
$\rho$, 0.005 for the scaling parameters  $\mathbf{S}$, and 0.001 for the rotation quaternions $\mathbf{R}$. The learning rate for the motion-model optimizer is set to 0.001. We trained 20,000 iterations and stopped. With the current implementation, the average reconstruction time for each sample was 475 minutes.

\section{Experiments and Results}

\subsection{Motion fields prediction}

\begin{figure}[htbp!]
\centering
   \includegraphics[width=1\textwidth]{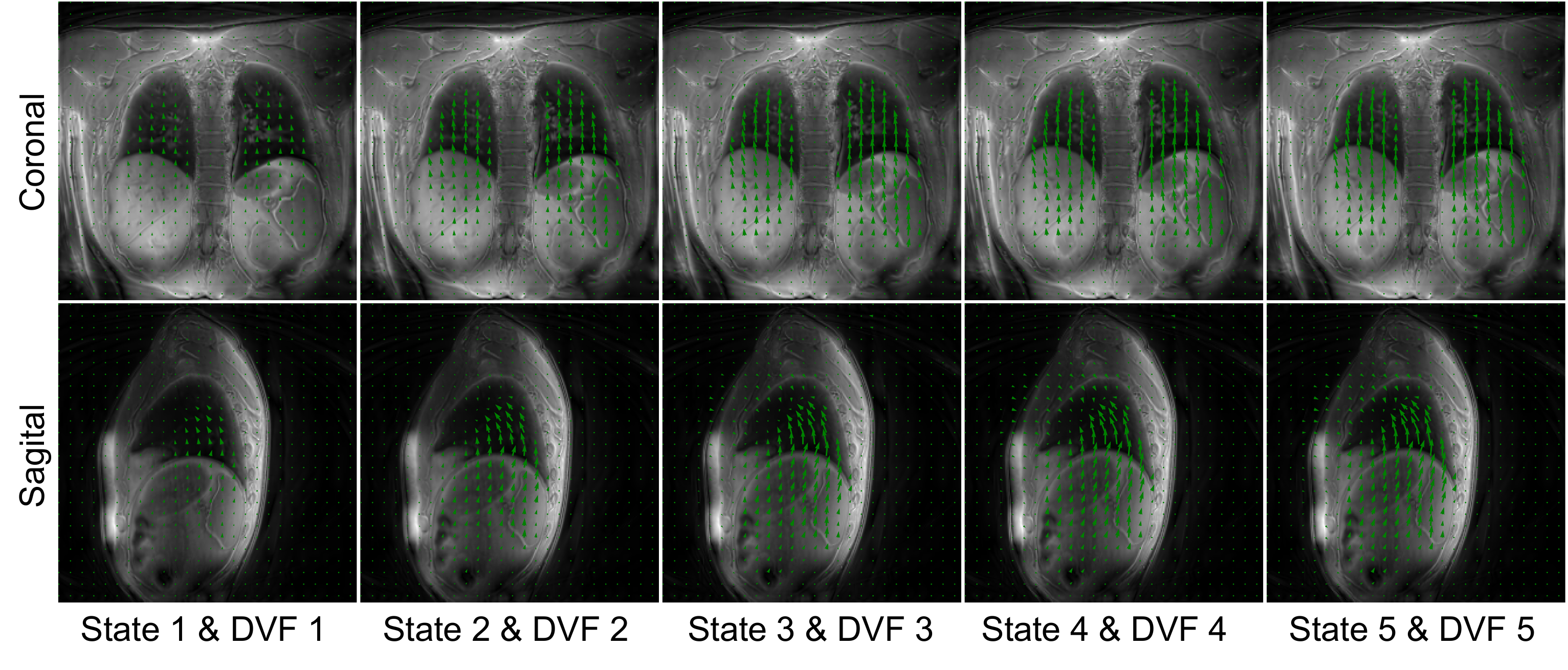}
  \caption{Examples of the predicted DVFs in the coronal and sagittal views. The DVFs from the first state to the subsequent motion states are overlaid on the corresponding current-state images.}
  \label{dvf}
\end{figure}

The proposed framework incorporates a learnable low-rank sparse motion model to directly estimate the DVFs from the first motion state to the subsequent states. In this model, the DIP-based CNN generates smooth and low-rank motion basis components, while the state-varying coefficients—one derived from the DC self-gating signal and the other learned adaptively—promote coherent motion variation across states. Figure \ref{dvf} presents representative DVFs in the sagittal and coronal planes across the 5 states when $N_{state}=5$ for a single slice, demonstrating the effectiveness of our motion model. As expected, the predicted motion fields exhibit coherent motion in the lungs and diaphragm regions, while the spinal column remains largely static, consistent with physiological behavior. These results further confirm the robustness of the proposed motion model in the motion-resolved scenario.

\subsection{Ablation study}
\begin{figure}[htbp!]
\centering
   \includegraphics[width=0.5\textwidth]{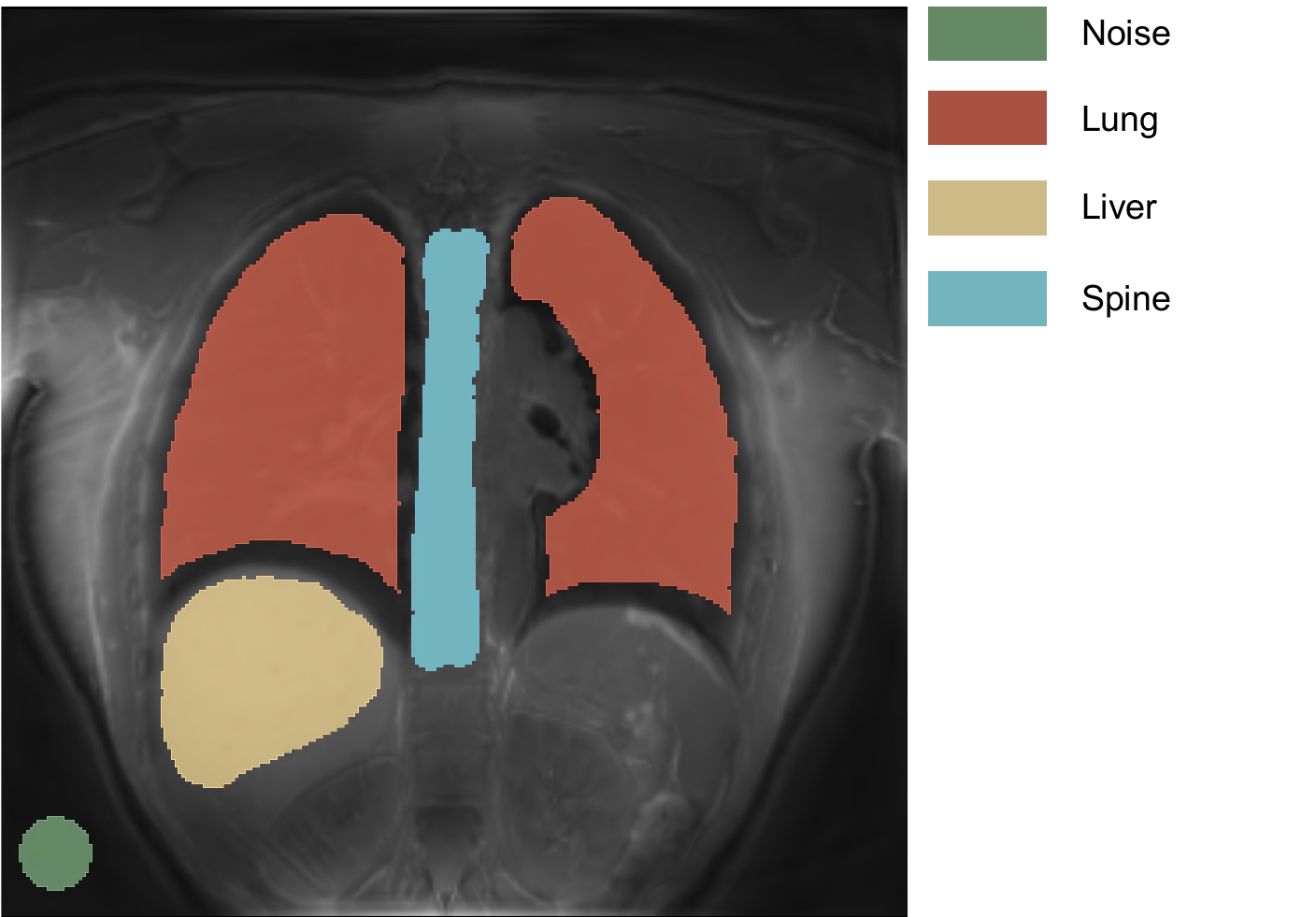}
  \caption{Representative coronal slice with the manually selected regions of interest (ROIs) used for quantitative evaluation. The ROIs correspond to the lung, liver, and spine regions, within which SNR and CNR were computed.}
  \label{coronal_masks}
\end{figure}

This section presents ablation studies on the MoRe-3DGSMR framework to assess the impact of three factors: the number of initialized Gaussian points, the TV regularization weight $\lambda_{TV}$, and the number of motion states used for k-space binning. In addition, we compared the proposed multi-resolution initialization strategy with the existing commonly used baseline methods: random initialization and equal-space initialization. The MoRe-3DGSMR model was trained in a self-supervised manner, without the use of ground truth.  To quantitatively assess the impact of different hyperparameter settings, several image quality metrics were evaluated on the reference frame of the reconstructed MR images, even though the proposed framework reconstructs $N_{state}$ different motion states. Quantitative metrics were computed on a selected coronal slice, where several representative anatomical regions, including the lung, liver, and spine, were manually defined as shown in Figure \ref{coronal_masks}. 

Signal-to-noise ratio (SNR) and contrast-to-noise ratio (CNR) are measured to analyze the image quality of the reconstructed images. SNR is calculated as follows:
\begin{equation}\label{key8}
SNR = 20 \log_{10} \left( \frac{\mu_{ROI}}{\sigma_{noise}} \right),    
\end{equation} 
where $\log_{10}$ is the base-10 logarithm, $\mu_{ROI}$ is the mean intensity value within the region of interest (ROI), and $\sigma_{noise}$ is the standard deviation of intensity values of the noise region that is chosen from the background of the images. Similarly, CNR assesses the contrast quality by comparing the difference in intensity values between the ROI and the selected noise regions through
\begin{equation}\label{key9}
CNR = 20 \log_{10} \left( \frac{|\mu_{ROI} - \mu_{noise}|}{\sigma_{noise}} \right),
\end{equation} 
where $\mu_{ROI}$ and $\mu_{noise}$ represent the mean intensity value of ROI and the noise regions while $\sigma_{noise}$ represents the standard deviation of the intensity values in the noise regions. SNR and CNR measure the difference between the selected ROI and the noise in the backgrounds. 

These quantitative metrics were computed on a selected coronal slice and are therefore slice-based; volumetric conclusions are further supported by consistent visual findings across coronal, sagittal, and axial views.

\subsubsection{Number of Gaussian points}

\begin{figure}[htbp!]
\centering
   \includegraphics[width=1\textwidth]{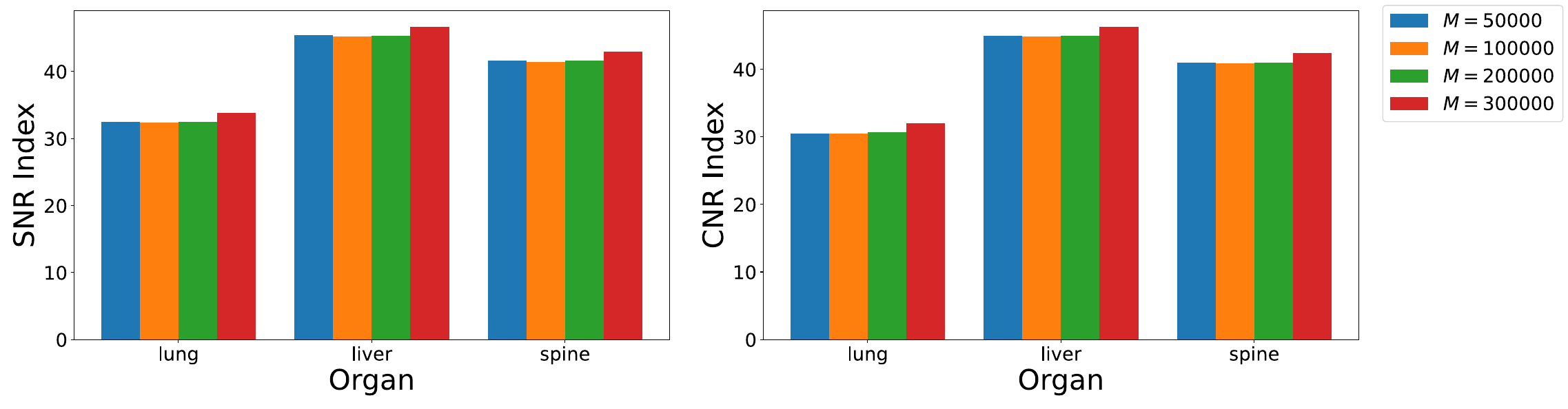}
  \caption{Results of the ablation study evaluating the effect of different numbers of Gaussian points are presented. Quantitative evaluation was performed on a selected coronal slice using SNR and CNR across three anatomical regions: the lung, liver, and spine.}
  \label{ablation_n_gaussians}
\end{figure}

\begin{figure}[htbp!]
\centering
   \includegraphics[width=1\textwidth]{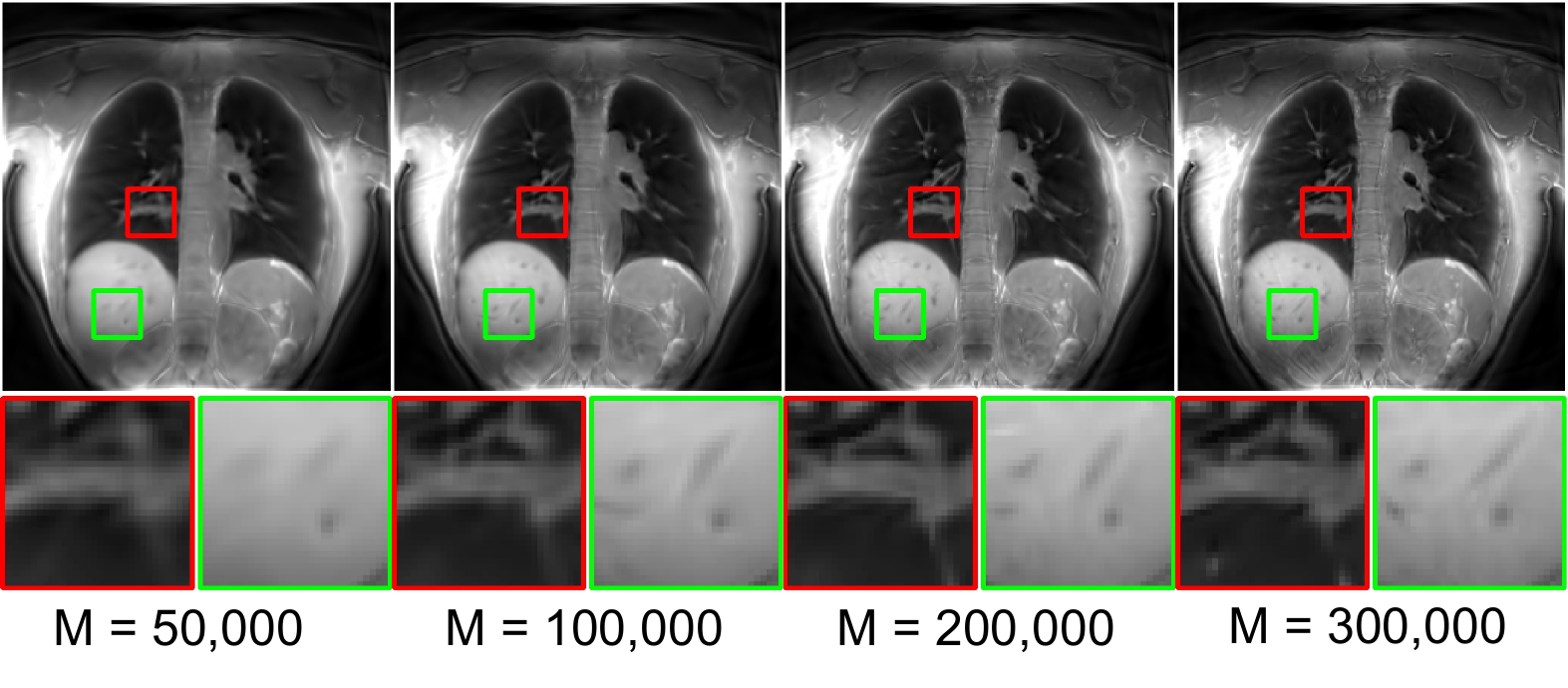}
  \caption{Visual comparisons from the ablation study evaluating the number of Gaussian point M are presented. The images showed the reference frame. }
  \label{n_gaussians_compare}
\end{figure}

 An ablation study was first performed to investigate the effect of the number of Gaussian points $M$ on reconstruction performance, using a fixed number of motion states ($N_{state}=6$) for a representative subject. All experiments were trained for 2,000 iterations using the proposed multi-resolution initialization strategy, without applying densification. To quantitatively assess image quality under different settings, we computed SNR and CNR. The results are presented in Figure \ref{ablation_n_gaussians}. The experimental results in Figure \ref{ablation_n_gaussians} indicate that,  $M = 300,000$ Gaussian points yields the best contrast, as evidenced by higher SNR and CNR. The visual comparisons are presented in Figure \ref{n_gaussians_compare}. Based on a comprehensive evaluation, we selected $M=300,000$ Gaussian point as the optimal configuration for the proposed framework.
 
\subsubsection{$\lambda_{TV}$ for TV regularization}
\begin{figure}[htbp!]
\centering
   \includegraphics[width=1\textwidth]{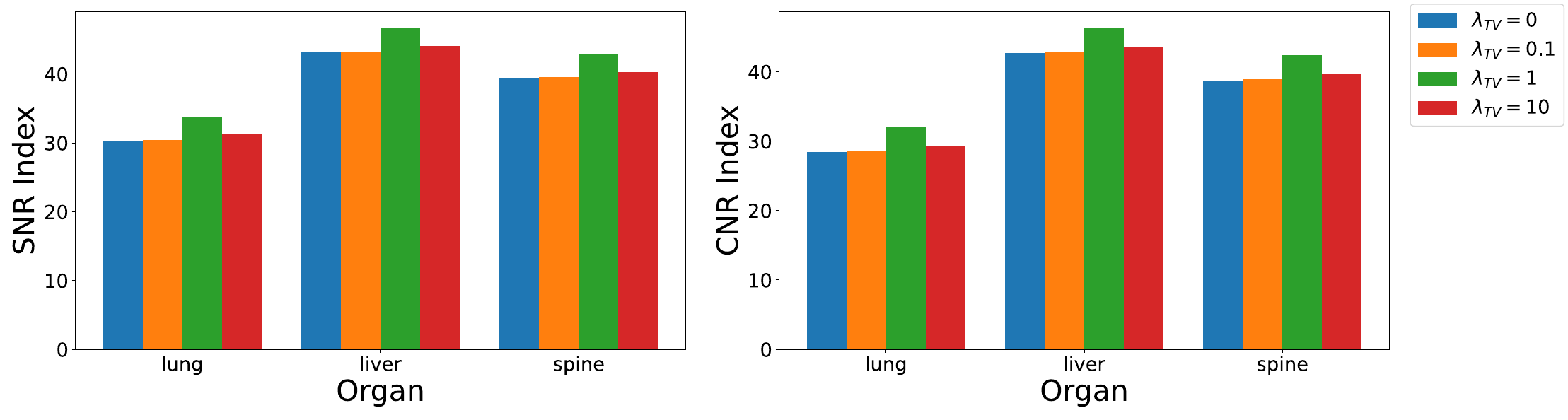}
  \caption{Results of the ablation study evaluating the effect of the hyperparameter $\lambda_{TV}$ are presented. Quantitative evaluation was performed on a selected coronal slice
using SNR and CNR, across
three anatomical regions: the lung, liver, and spine.}
  \label{tv_lambda}
\end{figure}

\begin{figure}[htbp!]
\centering
   \includegraphics[width=1\textwidth]{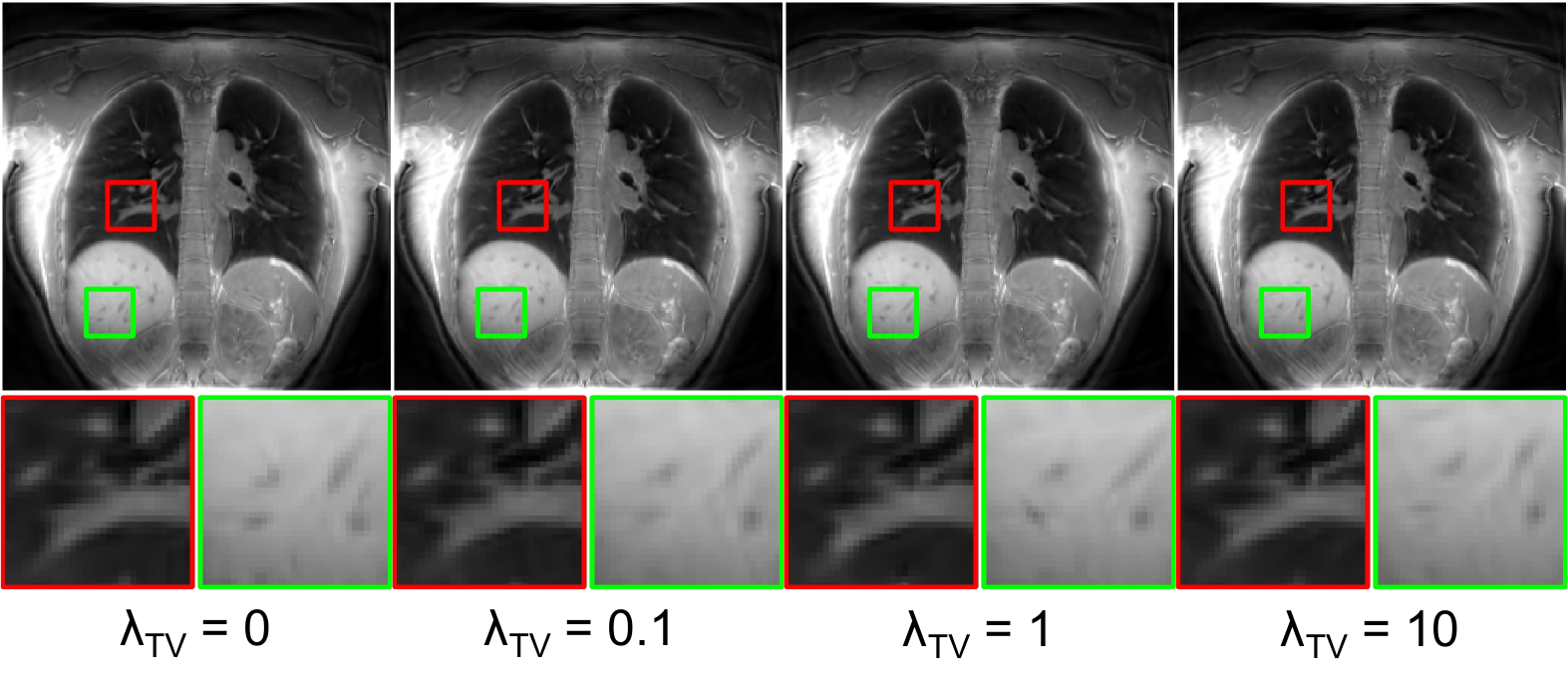}
  \caption{Visual comparisons from the ablation study evaluating $\lambda_{TV}$ are presented. The images showed the reference frame. }
  \label{tv_compare}
\end{figure}

An ablation study was conducted to investigate the impact of the TV regularization balancing parameter $\lambda_{TV}$ on image quality. For this analysis, the number of initialized Gaussian points was fixed at $M = 300,000$. The results are presented in Figure \ref{tv_lambda}. As shown, $\lambda_{TV}=1$ yields improved contrast, accompanied by higher SNR and CNR. Further increasing $\lambda_{TV}$ begins to reduce the contrast. This behavior is consistent with the expected trade-off of TV regularization, which suppresses noise while potentially compromising edge preservation. The visual comparisons are presented in Figure \ref{tv_compare}. Visual inspection further shows that $\lambda_{TV}=1$ provides a favorable balance between artifact suppression and structural preservation. In particular, the edges of the intrahepatic vessels in the liver region appear clearer and better defined, while background noise and residual artifacts are reduced compared with $\lambda_{TV}=0$ and $\lambda_{TV}=0.1$. When $\lambda_{TV}=10$, the image becomes more over-smoothed, and some fine vascular boundaries in the liver region become less distinct. Based on a comprehensive evaluation, $\lambda_{tv}=1$ was selected for this study.

\subsubsection{Effects of number of motion states $N_{state}$}
\begin{figure}[htbp!]
\centering
   \includegraphics[width=1\textwidth]{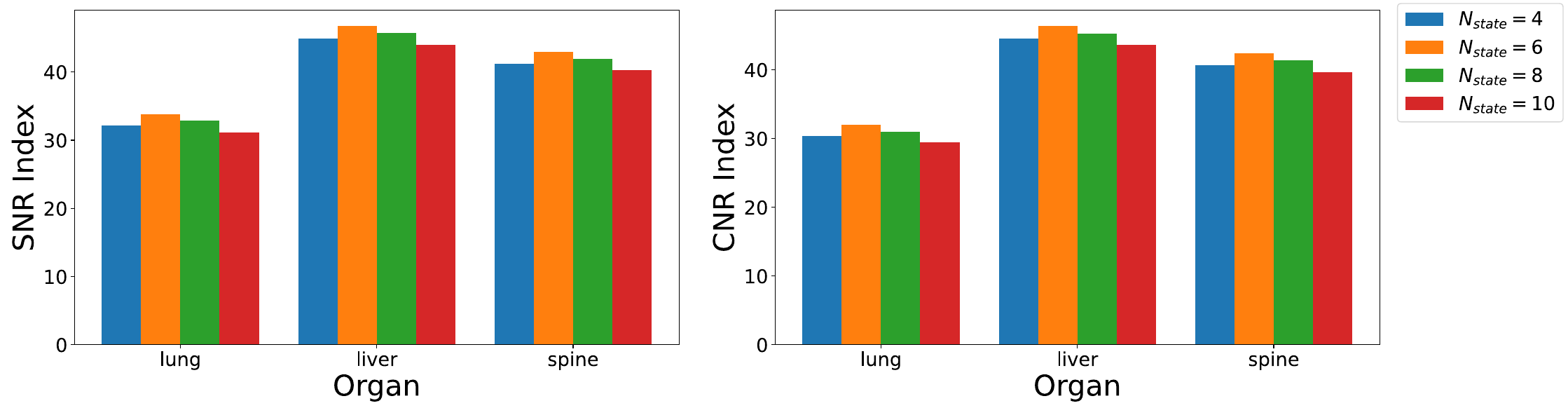}
  \caption{Results of the ablation study evaluating the effect of the number of motion states $N_{state}$ are presented. Quantitative assessments were performed using two evaluation metrics: SNR and CNR -- across three anatomical regions: lung, liver, and spine.}
  \label{motion_states1}
\end{figure}

\begin{figure}[htbp!]
\centering
   \includegraphics[width=1\textwidth]{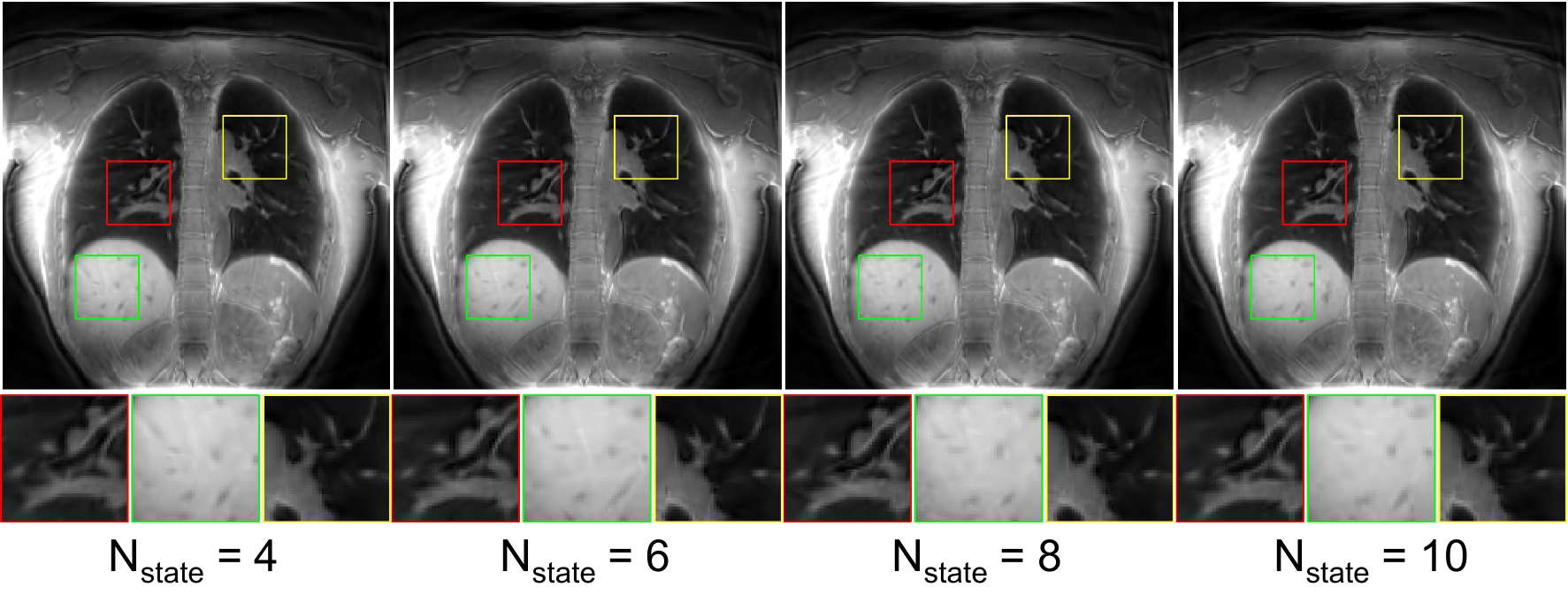}
  \caption{Visual comparison of the ablation study examining the number of motion-resolved states ($N_{state}$) is presented. The reconstructed reference state is included for direct comparison. Selected regions are zoomed in to facilitate detailed evaluation of structural preservation and image quality across different motion state configurations.}
\label{motion_states2}
\end{figure}

\begin{figure}[htbp!]
\centering
   \includegraphics[width=1\textwidth]{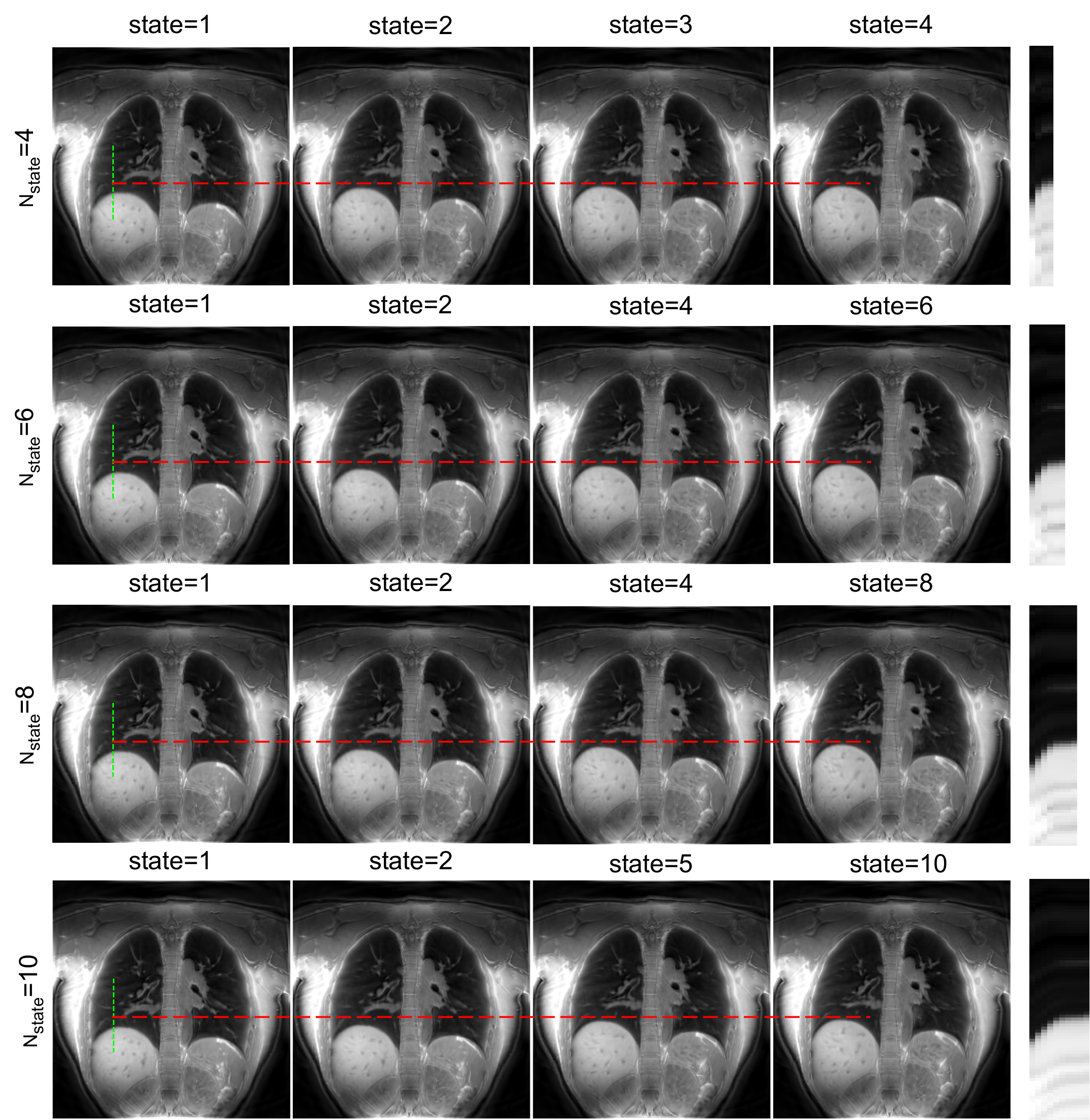}
  \caption{Visual comparisons from the ablation study evaluating the number of motion states ($N_{state}$) are presented. In each row, four reconstructed motion states corresponding to different phases of the respiratory cycle—end-inspiration, mid-inspiration, mid-expiration, and end-expiration—are shown to illustrate the temporal consistency and image quality across respiratory phases. A vertical line was placed through the superior aspect of the diaphragm, and the intensity values sampled along this line were stacked over states to generate the spatiotemporal profile shown on the right.}
  \label{motion_states3}
\end{figure}

In the proposed motion-resolved reconstruction framework, the acquired k-space data are retrospectively and uniformly sorted into $N_{state}$ respiratory motion states based on the estimated motion signal. This section investigates the influence of the number of motion states, $N_{state}$, on the quality of the reconstructed images. Different choices of $N_{state}$ result in different numbers of k-space radial spokes per state, which changes the amount of data available for reconstruction and consequently affects the level of recoverable image details. Quantitative results are presented in Figure \ref{motion_states1}. As shown, the configuration with $N_{state} = 6$ achieves the highest SNR and CNR. A visual comparison of reconstruction quality for different $N_{state}$ values is provided in Figure \ref{motion_states2}, focusing on the first motion state and including zoomed-in regions for detailed inspection. Additionally, representative images across different motion states for varying $N_{state}$ values are shown in Figure \ref{motion_states3} to further illustrate the effect of motion binning granularity. A vertical line was placed through the superior aspect of the diaphragm,
and the intensity values sampled along this line were stacked over states to generate the
spatiotemporal profile to present the motion of the diaphragm. Therefore, $N_{state} = 6$ was selected as the optimal setting, offering a favorable trade-off between motion resolution and noise performance.

\subsubsection{Multi-resolution initialization strategy for MoRe-3DGSMR} 
\begin{figure}[htbp!]
\centering
   \includegraphics[width=1\textwidth]{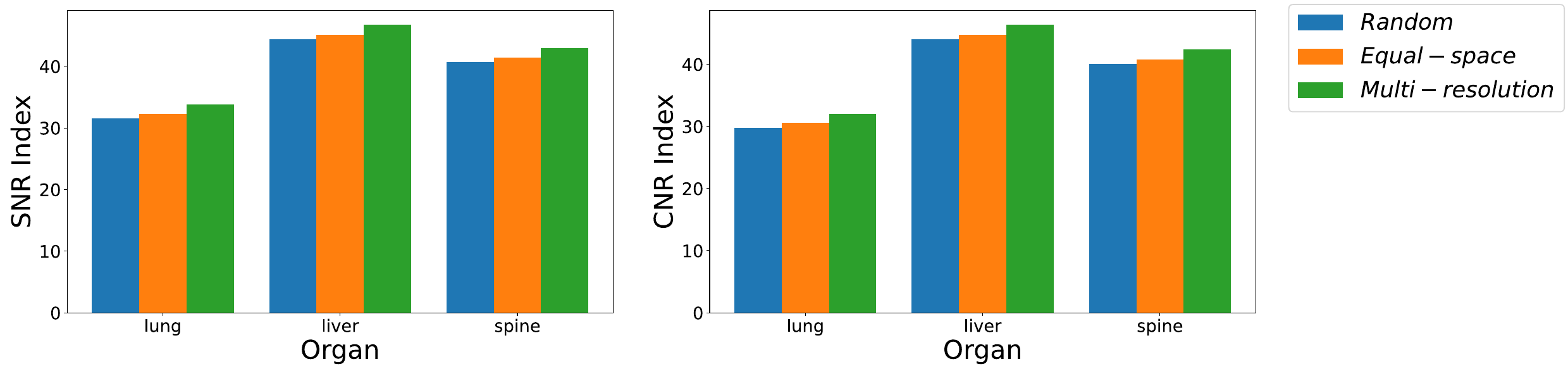}
  \caption{Results of the ablation study evaluating the effect of different 3DGS initialization strategies are presented. Quantitative evaluation was performed on a selected coronal slice using SNR and CNR, across three anatomical regions: the lung, liver, and spine.}
  \label{ablation_initialization}
\end{figure}

\begin{figure}[htbp!]
\centering
   \includegraphics[width=0.8\textwidth]{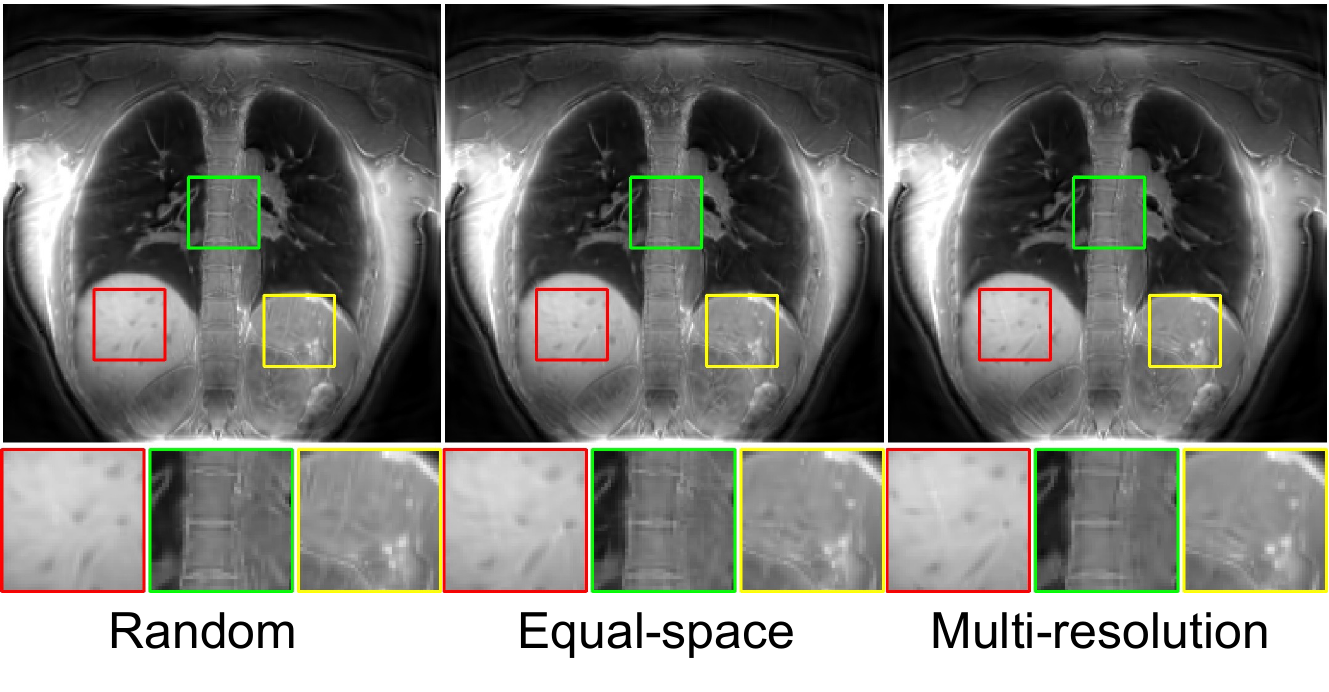}
  \caption{Visual comparisons from the ablation study on initialization strategies are presented. Detailed evaluations are performed on the reconstructed reference state, with selected regions zoomed in to highlight differences in structural preservation and image quality.}
  \label{init_qual}
\end{figure}

In this work, a multi-resolution initialization strategy was introduced within the MoRe-3DGSMR framework to enhance reconstruction performance. The effectiveness of this strategy was evaluated through both quantitative metrics (Figure \ref{ablation_initialization}) and qualitative visual comparisons (Figure \ref{init_qual}). The results consistently demonstrate the superiority of the proposed initialization approach. Specifically, Figure \ref{ablation_initialization} shows that the multi-resolution strategy yields improvements in SNR and CNR, relative to other commonly used initialization methods. Furthermore, the visual assessment presented in Figure \ref{init_qual} reveals enhanced preservation of fine anatomical details when using the multi-resolution initialization, further validating its advantage for high-fidelity reconstruction.

\begin{figure}[tb!]
\centering
   \includegraphics[width=\textwidth]{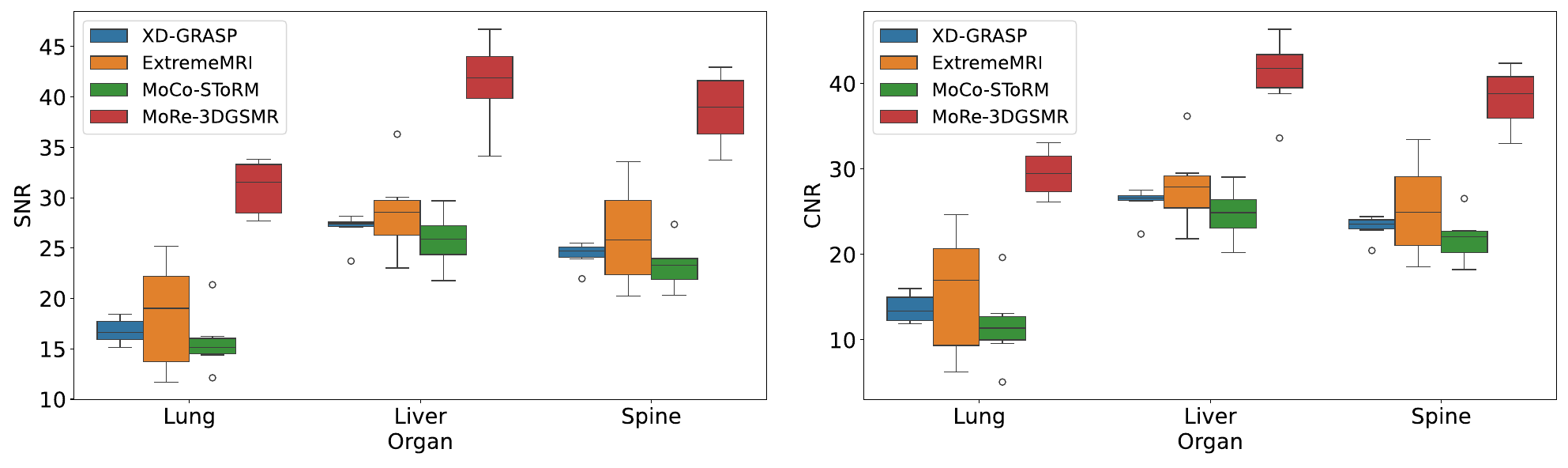}
  \caption{Quantitative comparisons of different reconstruction methods in lungs, livers, hearts, and spines: SNR and CNR.}
  \label{sota_quan}
\end{figure}

\begin{figure}[htbp!]
\centering
\includegraphics[height=1\textwidth]{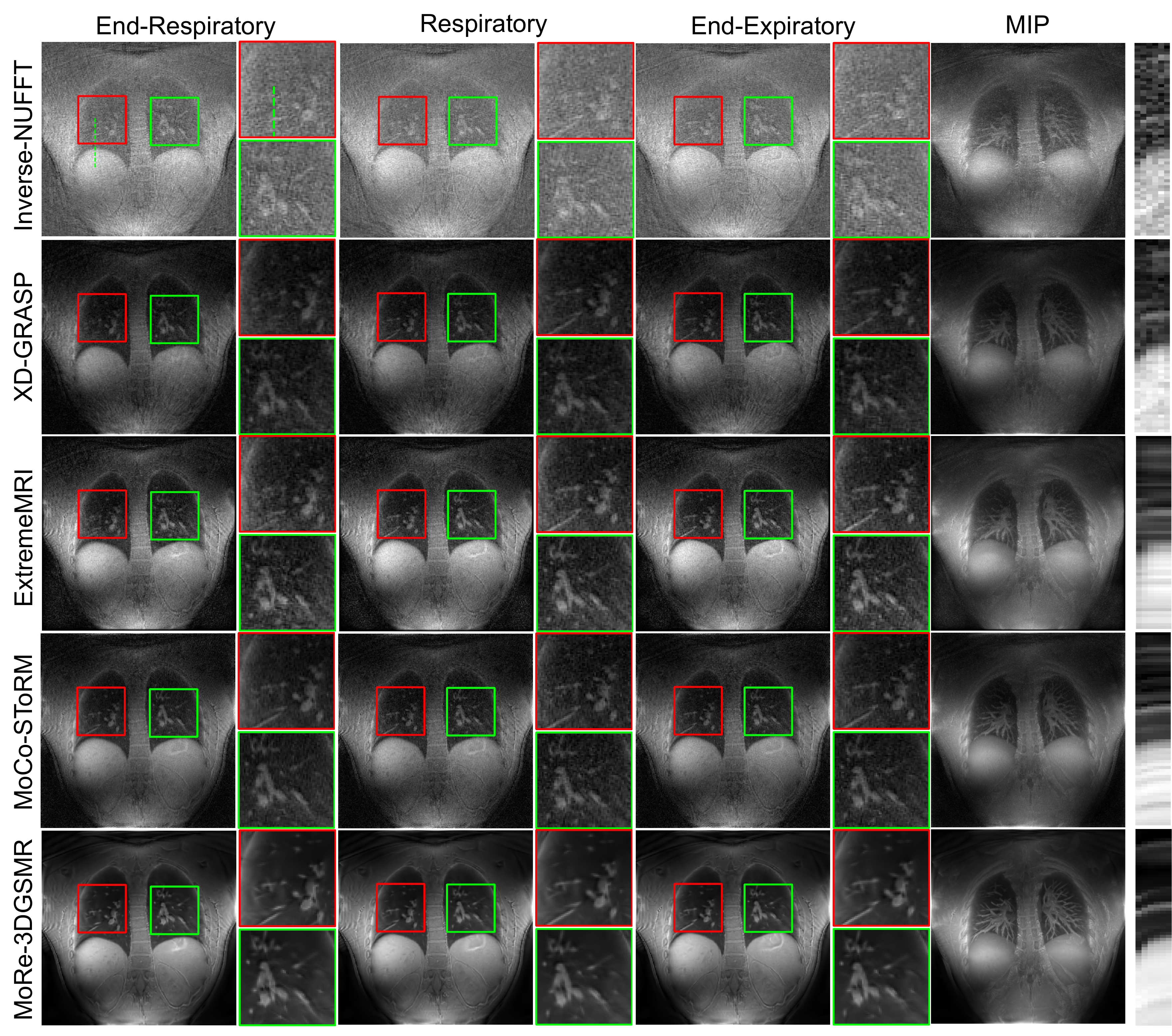}
  \caption{Visual comparisons between the proposed method and competing reconstruction approaches are presented. For each method, coronal slices corresponding to four respiratory phases—end-inspiration, mid-inspiration, mid-expiration, and end-expiration—are shown. Zoomed-in regions highlight the superior image quality achieved by the proposed method, particularly in preserving fine anatomical details. Additionally, maximum intensity projection (MIP) reconstructions of the central 20 slices are provided for all methods to further illustrate overall structural clarity and contrast.  A vertical line was placed through the superior aspect of the diaphragm, marked in green shown in the upper left corner image, and the intensity values sampled along this line were stacked over states to generate the
spatiotemporal profile shown on the right.}
  \label{sota_qual}
\end{figure}

\begin{figure}[tb!]
\centering
   \includegraphics[width=1\textwidth]{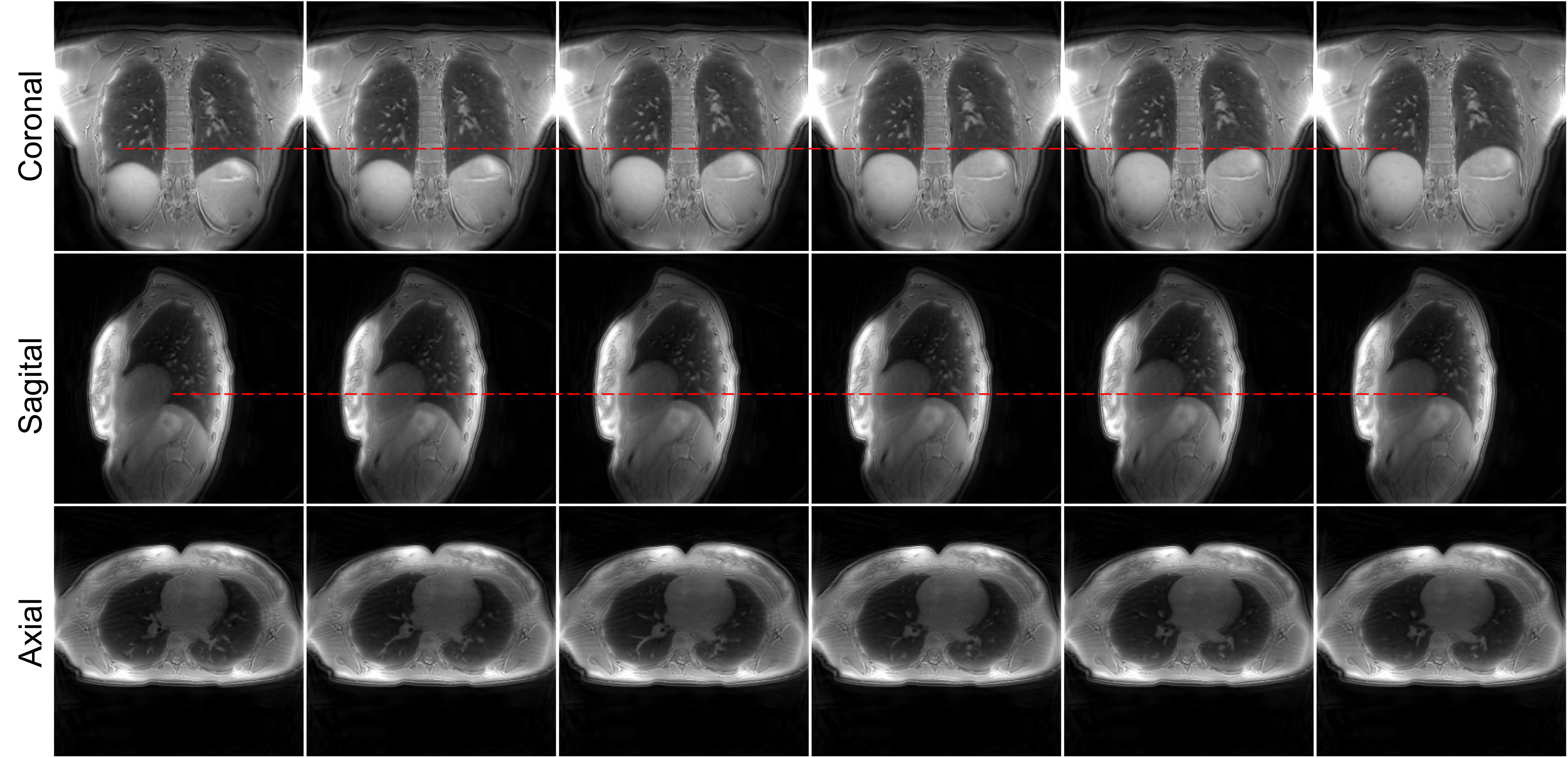}
  \caption{Showcase of reconstruction results on a representative subject in coronal, sagittal, and axial planes. }
  \label{showcase}
\end{figure}

\subsection{Comparison with the state-of-the-art methods}

In this section, the proposed MoRe-3DGSMR reconstruction framework is evaluated against three state-of-the-art dynamic MRI reconstruction methods: XD-GRASP, ExtremeMRI, and MoCo-SToRM. Quantitative comparisons were conducted using SNR and CNR metrics across four anatomical regions: lung, liver, heart, and spine. The results are summarized in Figure \ref{sota_quan}, where the boxplots indicate that the proposed method consistently outperforms the baseline methods in terms of both SNR and CNR.

Additionally, a qualitative comparison for a representative subject is presented in Figure \ref{sota_qual}. The visual results demonstrate that MoRe-3DGSMR produces reconstructions with reduced noise levels and improved depiction of fine structural details. 
% Regions with noticeable enhancements are highlighted using red and green arrows, further underscoring the superior visual fidelity of the proposed framework relative to existing techniques.

\subsection{Showcase of MoRe-3DGSMR}

This section presents representative reconstructed images from four respiratory motion states in three anatomical orientations for a single subject, as shown in Figure \ref{showcase}. Respiratory motion is visibly captured through the displacement of the diaphragm across states. Importantly, the reconstructed image quality remains consistently high across all orientations and motion states. Pulmonary vasculature is clearly delineated with excellent visibility, demonstrating the framework’s ability to preserve fine structural details throughout the respiratory cycle.

\subsection{Expand into time-resolved scenario}
\begin{figure}[tb!]
\centering
   \includegraphics[width=1\textwidth]{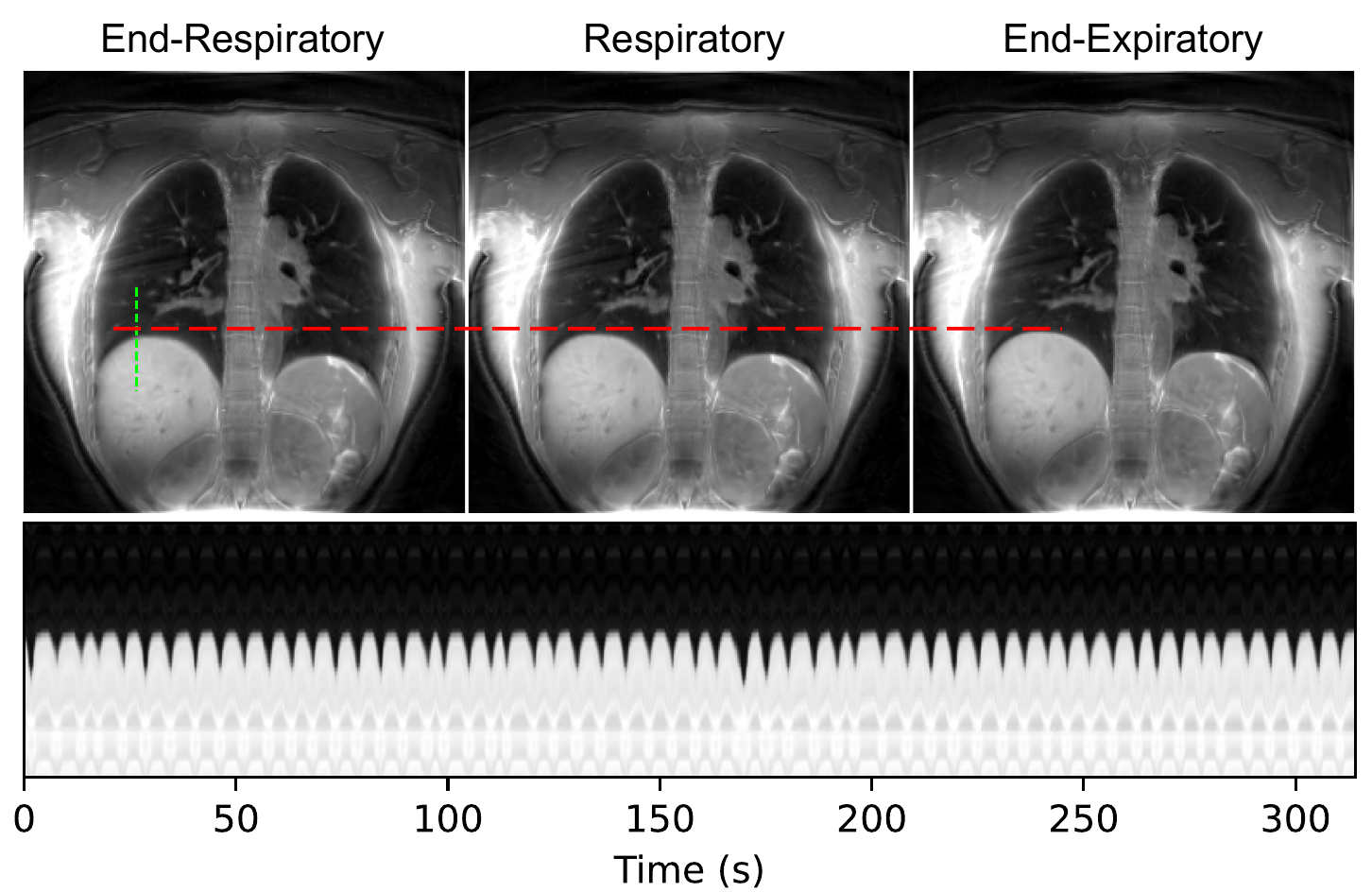}
  \caption{A representative image from a selected frame is shown here. A vertical line was positioned through the superior aspect of the diaphragm, and the intensity values sampled along this line were stacked across time to form the temporal profile.  }
  \label{time-resolved}
\end{figure}

Although the present work is formulated in a motion-resolved setting with discrete respiratory states, the proposed framework could be naturally extended to a time-resolved formulation by replacing the current discrete k-space binning, which is based on the derived DC self-gated respiratory signal, with a continuous temporal coordinate. Although this is not the main focus of the present study, we nevertheless conducted a preliminary feasibility experiment to demonstrate the potential of this extension. In this experiment, every 100 spokes were grouped into a single frame, corresponding to a temporal resolution of $100 \times TR = 0.37$ s per frame. The hyperparameters used in the model and loss function were kept nearly the same as in the motion-resolved setting. The temporal variation regularization was imposed on the estimated DVFs across two consecutive frames instead of across $N_{state}$ states. The results are shown in Figure \ref{time-resolved}.

\section{Discussion}

This study proposes an innovative motion-resolved reconstruction framework for 3D isotropic pulmonary MRI using a 3DGS framework, leveraging its effectiveness in adaptive smoothing and interpolation, as well as its powerful spatial representation, to reduce motion and undersampling artifacts. Based on the estimated motion signal from the center of k-space, the radial k-space data are divided into six motion states, each containing an equal number of radial k-space spokes. Although this binning strategy may introduce some motion artifacts because respiration is not perfectly periodic in practice, the smoothing capability of 3DGS may help mitigate these motion inconsistencies and thereby reduce the resulting artifacts. Our proposed learnable low-rank sparse motion model further helps reduce motion artifacts in the motion-resolved setting. In this model, each DVF is generated by combining two shared motion basis components with time-varying coefficients. One coefficient is derived from the median motion signal of each binned state, while the other is learned adaptively to refine the motion representation. As a result, the estimated DVFs are explicitly linked to the motion characteristics of the corresponding binned k-space data.

Since the Fourier transform of a Gaussian distribution remains Gaussian, the Fourier transform of MR images voxelized from a combination of 3D Gaussian points should also be represented as a combination of corresponding 3D Gaussian points in the frequency domain. Therefore, a well-designed undersampling scheme in the frequency domain can still effectively estimate these Gaussian point combinations. This is a key strength of 3DGS in MRI reconstruction. This study further confirms that this paradigm remains effective for non-Cartesian radial MRI reconstruction, where combinations of Gaussian distributions in the frequency domain can still be estimated to fit undersampled non-Cartesian radial k-space data. It is well known that NUFFT-based optimization can impose a substantial computational burden in motion-aware MRI reconstruction. As discussed in \cite{peng2025three}, a Gaussian representation of complex-valued MRI has the attractive property that it can, in principle, be projected directly into the frequency domain, because the Fourier transform of a Gaussian distribution remains Gaussian. This offers the potential to avoid repeated NUFFT operations and thereby substantially accelerate training. In the present study, we focused on the motion-resolved setting rather than the time-resolved setting. One important reason is computational practicality: in a time-resolved formulation, the number of reconstructed frames increases substantially, and the associated NUFFT-based forward and backward operations would lead to a dramatic increase in training time and memory costs. In contrast, motion-resolved reconstruction groups the data into a limited number of respiratory states, which provides a more balanced amount of k-space data for each state and makes the training process more tractable. This setting also tends to yield better spatial image quality, since each reconstructed state is supported by a relatively uniform amount of k-space data.

The multi-resolution initialization strategy improves optimization stability and convergence by explicitly assigning Gaussian points of different scales to different k-space frequency bands and optimizing them jointly in a coarse-to-fine manner. Specifically, larger Gaussians predominantly fit low-frequency components (global structure), while smaller Gaussians capture high-frequency components (fine details), so the reconstruction is refined across frequency levels within a unified training process rather than staged fitting. However, its benefit depends on the availability of sufficient signal support in each motion state. Under highly irregular breathing patterns, state binning may become inconsistent across cycles, reducing motion coherence and weakening the advantage of structured initialization. In addition, under extremely sparse sampling per state, high-frequency components may be insufficiently constrained, and reconstruction can become over-smoothed with reduced visibility of subtle fine details. Therefore, the proposed framework is most suitable for free-breathing acquisitions with relatively stable respiratory periodicity and moderate sampling density per state.

A broader challenge in motion-resolved MRI is that both image quality and motion accuracy remain difficult to evaluate objectively. Although SNR and CNR were used in this study, these metrics depend on ROI selection and therefore cannot fully reflect overall reconstruction quality. In practice, evaluation still relies heavily on visual inspection, including artifact level, anatomical plausibility, and preservation of fine structures, making it inherently subjective. In addition, since ground-truth motion or deformation fields are generally unavailable, it is difficult to directly verify whether the estimated DVFs are fully accurate. This is particularly challenging for small vessels and subtle anatomical structures. Therefore, the estimated motion can only be assessed indirectly, mainly through the smoothness and coherence of the recovered motion, the reduction of motion artifacts, and the plausibility of visible motion patterns such as diaphragm displacement. This remains an open challenge in motion-resolved MRI.

The Gaussian-based representation introduces an inherent smoothness bias, which is both a strength and a limitation of 3DGS in motion-resolved MRI reconstruction. Under undersampled acquisition, high-frequency components are already more difficult to recover faithfully, and the continuous Gaussian representation further favors smooth and spatially coherent structures. This makes the framework naturally robust to undersampling and beneficial for suppressing noise, incoherent fluctuations, and motion-related artifacts, thereby stabilizing the reconstruction and preserving the dominant anatomical content. However, the same property may also lead to oversmoothing and reduce the visibility of fine high-frequency details, such as small vessels, thin boundaries, or subtle anatomical textures. Consequently, the reconstructed images may show relatively clear major anatomical edges together with smoother homogeneous tissue regions. This suggests that 3DGS is particularly effective for recovering the main anatomical structure under limited data, but may be less effective in faithfully preserving all fine-scale details.

\section{Conclusion}

In this work, we propose a motion-resolved reconstruction framework by extending 3DGS to 3D non-Cartesian radial pulmonary MRI reconstruction. The results outperform existing baseline reconstruction methods, highlighting the potential of the proposed framework. The proposed method reconstructs images with finer details across different structures and organs, further demonstrating its clinical potential for pulmonary MRI.

 \bibliographystyle{elsarticle-num} 
 \bibliography{refs}

\begin{thebibliography}{10}
\expandafter\ifx\csname url\endcsname\relax
  \def\url#1{\texttt{#1}}\fi
\expandafter\ifx\csname urlprefix\endcsname\relax\def\urlprefix{URL }\fi
\expandafter\ifx\csname href\endcsname\relax
  \def\href#1#2{#2} \def\path#1{#1}\fi

\bibitem{hatabu2020expanding}
H.~Hatabu, Y.~Ohno, W.~B. Gefter, G.~Parraga, B.~Madore, K.~S. Lee, T.~A. Altes, D.~A. Lynch, J.~R. Mayo, J.~B. Seo, et~al., Expanding applications of pulmonary mri in the clinical evaluation of lung disorders: Fleischner society position paper, Radiology 297~(2) (2020) 286--301.

\bibitem{ohno2022overview}
Y.~Ohno, S.~Hanamatsu, Y.~Obama, T.~Ueda, H.~Ikeda, H.~Hattori, K.~Murayama, H.~Toyama, Overview of mri for pulmonary functional imaging, The British Journal of Radiology 95~(1132) (2022) 20201053.

\bibitem{sharma2022quantification}
M.~Sharma, P.~V. Wyszkiewicz, V.~Desaigoudar, F.~Guo, D.~P. Capaldi, G.~Parraga, Quantification of pulmonary functional mri: state-of-the-art and emerging image processing methods and measurements, Physics in Medicine \& Biology 67~(22) (2022) 22TR01.

\bibitem{sanchez2023detection}
F.~Sanchez, P.~N. Tyrrell, P.~Cheung, C.~Heyn, S.~Graham, I.~Poon, Y.~Ung, A.~Louie, M.~Tsao, A.~Oikonomou, Detection of solid and subsolid pulmonary nodules with lung mri: performance of ute, t1 gradient-echo, and single-shot t2 fast spin echo, Cancer Imaging 23~(1) (2023) 17.

\bibitem{larson2024lung}
P.~Larson, Lung imaging with ute-mri, in: MRI of Short-and Ultrashort-T2 Tissues: Making the Invisible Visible, Springer, 2024, pp. 527--534.

\bibitem{wang2025assessment}
X.~Wang, Y.~Cui, Y.~Wang, S.~Liu, N.~Meng, W.~Wei, Y.~Bai, Y.~Shen, J.~Guo, Z.~Guo, et~al., Assessment of lung nodule detection and lung ct screening reporting and data system classification using zero echo time pulmonary mri, Journal of Magnetic Resonance Imaging 61~(2) (2025) 822--829.

\bibitem{ufuk2024comparing}
F.~Ufuk, B.~Kurnaz, H.~Peker, E.~Sagtas, Z.~D. Ok, V.~Cobankara, Comparing three-dimensional zero echo time (3d-zte) lung mri and chest ct in the evaluation of systemic sclerosis-related interstitial lung disease, European Radiology (2024) 1--10.

\bibitem{gandhi2024comparison}
D.~B. Gandhi, N.~S. Higano, A.~D. Hahn, C.~C. Gunatilaka, L.~A. Torres, S.~B. Fain, J.~C. Woods, A.~J. Bates, Comparison of weighting algorithms to mitigate respiratory motion in free-breathing neonatal pulmonary radial ute-mri, Biomedical physics \& engineering express 10~(3) (2024) 035030.

\bibitem{yu2021free}
N.~Yu, H.~Duan, C.~Yang, Y.~Yu, S.~Dang, Free-breathing radial 3d fat-suppressed t1-weighted gradient echo (r-vibe) sequence for assessment of pulmonary lesions: a prospective comparison of ct and mri, Cancer Imaging 21 (2021) 1--9.

\bibitem{wu20234d}
C.~Wu, G.~Krishnamoorthy, V.~Yu, E.~Subashi, A.~Rimner, R.~Otazo, 4d lung mri with high-isotropic-resolution using half-spoke (ute) and full-spoke 3d radial acquisition and temporal compressed sensing reconstruction, Physics in Medicine \& Biology 68~(3) (2023) 035017.

\bibitem{johnson2013optimized}
K.~M. Johnson, S.~B. Fain, M.~L. Schiebler, S.~Nagle, Optimized 3d ultrashort echo time pulmonary mri, Magnetic resonance in medicine 70~(5) (2013) 1241--1250.

\bibitem{zhu2021optimizing}
X.~Zhu, F.~Tan, K.~Johnson, P.~Larson, Optimizing trajectory ordering for fast radial ultra-short te (ute) acquisitions, Journal of Magnetic Resonance 327 (2021) 106977.

\bibitem{zou2023time}
Q.~Zou, Z.~Miller, S.~Dzelebdzic, M.~Abadeer, K.~M. Johnson, T.~Hussain, Time-resolved 3d cardiopulmonary mri reconstruction using spatial transformer network, Mathematical Biosciences and Engineering 20~(9) (2023) 15982--15998.

\bibitem{zou2024motion}
Q.~Zou, Motion-resolved 3d pulmonary mri reconstruction using sinusoidal representation networks, Current Medical Imaging 20~(1) (2024) e15734056270732.

\bibitem{montazeri2021design}
K.~Montazeri, S.~A. Jonsson, J.~S. Agustsson, M.~Serwatko, T.~Gislason, E.~S. Arnardottir, The design of rip belts impacts the reliability and quality of the measured respiratory signals, Sleep and Breathing (2021) 1--7.

\bibitem{feng2014golden}
L.~Feng, R.~Grimm, K.~T. Block, H.~Chandarana, S.~Kim, J.~Xu, L.~Axel, D.~K. Sodickson, R.~Otazo, Golden-angle radial sparse parallel mri: combination of compressed sensing, parallel imaging, and golden-angle radial sampling for fast and flexible dynamic volumetric mri, Magnetic resonance in medicine 72~(3) (2014) 707--717.

\bibitem{feng2016xd}
L.~Feng, L.~Axel, H.~Chandarana, K.~T. Block, D.~K. Sodickson, R.~Otazo, Xd-grasp: golden-angle radial mri with reconstruction of extra motion-state dimensions using compressed sensing, Magnetic resonance in medicine 75~(2) (2016) 775--788.

\bibitem{ong2020extreme}
F.~Ong, X.~Zhu, J.~Y. Cheng, K.~M. Johnson, P.~E. Larson, S.~S. Vasanawala, M.~Lustig, Extreme mri: Large-scale volumetric dynamic imaging from continuous non-gated acquisitions, Magnetic resonance in medicine 84~(4) (2020) 1763--1780.

\bibitem{murray2024movienet}
V.~Murray, S.~Siddiq, C.~Crane, M.~El~Homsi, T.-H. Kim, C.~Wu, R.~Otazo, Movienet: deep space--time-coil reconstruction network without k-space data consistency for fast motion-resolved 4d mri, Magnetic resonance in medicine 91~(2) (2024) 600--614.

\bibitem{jafari2023graspnet}
R.~Jafari, R.~K.~G. Do, M.~D. LaGratta, M.~Fung, E.~Bayram, T.~Cashen, R.~Otazo, Graspnet: fast spatiotemporal deep learning reconstruction of golden-angle radial data for free-breathing dynamic contrast-enhanced magnetic resonance imaging, NMR in Biomedicine 36~(3) (2023) e4861.

\bibitem{zou2022dynamic}
Q.~Zou, L.~A. Torres, S.~B. Fain, N.~S. Higano, A.~J. Bates, M.~Jacob, Dynamic imaging using motion-compensated smoothness regularization on manifolds (moco-storm), Physics in medicine \& biology 67~(14) (2022) 144001.

\bibitem{li2018efficient}
K.~Li, T.~Pham, H.~Zhan, I.~Reid, Efficient dense point cloud object reconstruction using deformation vector fields, in: Proceedings of the European Conference on Computer Vision (ECCV), 2018, pp. 497--513.

\bibitem{kerbl20233d}
B.~Kerbl, G.~Kopanas, T.~Leimk{\"u}hler, G.~Drettakis, 3d gaussian splatting for real-time radiance field rendering., ACM Trans. Graph. 42~(4) (2023) 139--1.

\bibitem{bersillon2001use}
J.-L. Bersillon, F.~Villieras, F.~Bardot, T.~Gorner, J.-M. Cases, Use of the gaussian distribution function as a tool to estimate continuous heterogeneity in adsorbing systems, Journal of colloid and interface science 240~(2) (2001) 400--411.

\bibitem{zha2024r}
R.~Zha, T.~J. Lin, Y.~Cai, J.~Cao, Y.~Zhang, H.~Li, R $r^2$-gaussian: Rectifying radiative gaussian splatting for tomographic reconstruction, arXiv preprint arXiv:2405.20693 (2024).

\bibitem{peng2025three}
T.~Peng, R.~Zha, Z.~Li, X.~Liu, Q.~Zou, Three-dimensional mri reconstruction with 3d gaussian representations: Tackling the undersampling problem, IEEE Transactions on Medical Imaging (2025).

\bibitem{li20113d}
R.~Li, J.~H. Lewis, X.~Jia, X.~Gu, M.~Folkerts, C.~Men, W.~Y. Song, S.~B. Jiang, 3d tumor localization through real-time volumetric x-ray imaging for lung cancer radiotherapy, Medical physics 38~(5) (2011) 2783--2794.

\bibitem{li2011pca}
R.~Li, J.~H. Lewis, X.~Jia, T.~Zhao, W.~Liu, S.~Wuenschel, J.~Lamb, D.~Yang, D.~A. Low, S.~B. Jiang, On a pca-based lung motion model, Physics in Medicine \& Biology 56~(18) (2011) 6009--6030.

\bibitem{shao2025real}
H.-C. Shao, T.~Mengke, T.~Pan, Y.~Zhang, Real-time cbct imaging and motion tracking via a single arbitrarily-angled x-ray projection by a joint dynamic reconstruction and motion estimation (dreme) framework, Physics in Medicine \& Biology 70~(2) (2025) 025026.

\bibitem{huang2025digs}
Y.~Huang, I.~Singh, T.~Joyce, K.~Thielemans, J.~R. McClelland, Digs: Dynamic cbct reconstruction using deformation-informed 4d gaussian splatting and a low-rank free-form deformation model, in: International Conference on Medical Image Computing and Computer-Assisted Intervention, Springer, 2025, pp. 131--141.

\bibitem{chen2025multi}
C.~Chen, M.~Vornehm, Z.~Bu, P.~Chandrasekaran, M.~A. Sultan, S.~M. Arshad, Y.~Liu, Y.~Han, R.~Ahmad, A multi-dynamic low-rank deep image prior (ml-dip) for 3d real-time cardiovascular mri, Journal of Cardiovascular Magnetic Resonance (2025) 102015.

\bibitem{liang2025analysis}
S.~Liang, E.~Bell, Q.~Qu, R.~Wang, S.~Ravishankar, Analysis of deep image prior and exploiting self-guidance for image reconstruction, IEEE Transactions on Computational Imaging (2025).

\bibitem{zhu2020iterative}
X.~Zhu, M.~Chan, M.~Lustig, K.~M. Johnson, P.~E. Larson, Iterative motion-compensation reconstruction ultra-short te (imoco ute) for high-resolution free-breathing pulmonary mri, Magnetic resonance in medicine 83~(4) (2020) 1208--1221.

\bibitem{zhang2016robust}
T.~Zhang, J.~Y. Cheng, Y.~Chen, D.~G. Nishimura, J.~M. Pauly, S.~S. Vasanawala, Robust self-navigated body mri using dense coil arrays, Magnetic resonance in medicine 76~(1) (2016) 197--205.

\bibitem{rodriguez2013total}
P.~Rodr{\'\i}guez, Total variation regularization algorithms for images corrupted with different noise models: a review, Journal of Electrical and Computer Engineering 2013~(1) (2013) 217021.

\bibitem{charatan2024pixelsplat}
D.~Charatan, S.~L. Li, A.~Tagliasacchi, V.~Sitzmann, pixelsplat: 3d gaussian splats from image pairs for scalable generalizable 3d reconstruction, in: Proceedings of the IEEE/CVF conference on computer vision and pattern recognition, 2024, pp. 19457--19467.

\bibitem{jung2024relaxing}
J.~Jung, J.~Han, H.~An, J.~Kang, S.~Park, S.~Kim, Relaxing accurate initialization constraint for 3d gaussian splatting, arXiv preprint arXiv:2403.09413 (2024).

\bibitem{pedersen2009k}
H.~Pedersen, S.~Kozerke, S.~Ringgaard, K.~Nehrke, W.~Y. Kim, k-t pca: temporally constrained k-t blast reconstruction using principal component analysis, Magnetic Resonance in Medicine: An Official Journal of the International Society for Magnetic Resonance in Medicine 62~(3) (2009) 706--716.

\end{thebibliography}



\end{document}